\documentclass[a4paper,fleqn,usenatbib]{mnras}
\usepackage{aasmacros}
\usepackage{times}
\usepackage{amsmath}
\usepackage{graphicx}
\usepackage{amssymb}
\usepackage[usenames,dvipsnames]{color}
\usepackage{url}
\usepackage{fixltx2e}
\usepackage{mathtools}
\usepackage{amsmath}
\usepackage{float}
\usepackage{cite}
\usepackage{fixltx2e}
\bibpunct{(}{)}{;}{a}{,}{,}
\usepackage{longtable}
\usepackage{hyperref}
\usepackage{morefloats}
\usepackage{gensymb}
\usepackage{enumitem}
\usepackage{tabu}
\usepackage{totpages}
\usepackage{amssymb}
\usepackage{xcolor}
\usepackage[T1]{fontenc}
\usepackage{ae,aecompl}

\usepackage{ mathrsfs }

\let\cite=\citen

\def\apj{ApJ}
\def\aj{AJ}
\def\mnras{MNRAS}
\def\pasa{PASA}

\bibpunct{(}{)}{;}{a}{}{,}

\def\prl{Phys. Rev. (Lett.)}

\makeatletter
\newcommand*{\rom}[1]{\expandafter\@slowromancap\romannumeral #1@}
\makeatother

\title[Filament Stacking]{Discovery of Magnetic Fields Along Stacked Cosmic Filaments as Revealed by Radio and X-Ray Emission}

\author[Vernstrom et. al]{T. Vernstrom\thanks{E-mail:tessa.vernstrom@csiro.au}$^1$, Heald, G.$^{1}$, Vazza, F.$^{2,3,4}$, Galvin, T. J.$^{1,5}$, West, J.L.$^{6}$, Locatelli, N.$^{2,3}$,\newauthor
Fornengo, N.$^{7}$, Pinetti, E.$^{7,8}$ \\ 
$^1$CSIRO Astronomy $\&$ Space Science, Kensington, Perth 6151, Australia\\
$^2$Dipartimento di Fisica e Astronomia, Universita di Bologna, Via Gobetti 93/2, 40122, Bologna, Italy \\
$^3$INAF, Istituto di Radioastronomia di Bologna, via Gobetti 101, I-41029 Bologna, Italy\\
$^4$Hamburger Sternwarte, Gojenbergsweg 112, 21029 Hamburg, Germany \\
$^5$International Centre for Radio Astronomy Research, Curtin University, Bentley, WA 6102, Australia\\
$^6$Dunlap Institute for Astronomy and Astrophysics University of Toronto, Toronto, ON M5S 3H4, Canada\\
$^7$Dipartimento di Fisica, Universit\`{a} di Torino and INFN, Sezione di Torino, via P. Giuria 1, 10125, Torino, Italy\\
$^8$Sorbonne Universit\'e and Laboratoire de Physique Th\'eorique et Hautes Energies (LPTHE), UMR 7589 CNRS, 4 Place Jussieu, 75252, \\Paris, France
 }

\begin{document}
  
\pagerange{\pageref{firstpage}--\pageref{lastpage}} \pubyear{2019}

\maketitle

\label{firstpage}
\begin{abstract}
Diffuse filaments connect galaxy clusters to form the cosmic web. Detecting these filaments could yield information on the magnetic field strength, cosmic ray population and temperature of intercluster gas, yet, the faint and large-scale nature of these bridges makes direct detections very challenging. Using multiple independent all-sky radio and X-ray maps we stack pairs of luminous red galaxies as tracers for cluster pairs. For the first time, we detect an average surface brightness between the clusters from synchrotron (radio) and thermal (X-ray) emission with  $\gtrsim 5\sigma$ significance, on physical scales larger than observed to date ($\geq 3\,$Mpc). We obtain a synchrotron spectral index of $\alpha \simeq -1.0$ and estimates of the average magnetic field strength of $ 30 \leq B \leq 60 \,$nG, derived from both equipartition and Inverse Compton arguments, implying a 5 to 15$\,$per cent degree of field regularity when compared with Faraday rotation measure estimates. While the X-ray detection is inline with predictions, the average radio signal comes out higher than predicted by cosmological simulations and dark matter annihilation and decay models. This discovery demonstrates that there are connective structures between mass concentrations that are significantly magnetised, and the presence of sufficient cosmic rays to produce detectable synchrotron radiation. 
\end{abstract}


\section{Introduction}
\label{sec:introduction}

Galaxy groups and clusters are the largest gravitationally bound structures in the Universe, and are expected to be connected by diffuse filaments. The filaments connecting these clusters are thought to be mostly permeated by warm-hot ($T \simeq 10^5$ - $10^7\,$ K) diffuse gas, known as the warm hot intergalactic medium \citep[WHIM, e.g. ][]{Dave01,Gheller15}. However, many details of the physical nature of filaments remain a mystery. It is thought that magnetic fields also permeate these filamentary regions, and that strong accretion shocks from in-falling matter can accelerate particles to relativistic energies, producing synchrotron emission \citep{Keshet04,Skillman08}. However, we do not know the strength of magnetic fields in filaments, how turbulent they are, or the brightness of the expected synchrotron emission. Direct imaging of filaments has not been possible with current telescopes due to their faint and diffuse nature, except in a very small number of extreme cases.

There have been few direct detections of hot gas filaments in the X-ray \citep{Werner08,Eckert15}, as well as a recent detection via stacking \citep{Tanimura20b}. The most recent direct detection from \citet{Reiprich20}, used the extended ROentgen Survey with an Imaging Telescope Array (eROSITA) telescope to measure a $15\,$Mpc long X-ray filament between two clusters. These studies suggest temperatures of order $10^5$-$10^7 \,$K, densities of $10$-100 times the average cosmic value and about 5-10$\,$per cent of their mass in baryonic gas. These works demonstrate the existence of an energetic large-scale baryon population. 

There have been a number of recent works in the radio to understand the synchrotron and magnetic field properties of the cosmic web. \citet{Vernstrom17} and \citet{Brown17} used cross-correlation analysis to obtain upper limits for the intergalactic magnetic field (IGMF) strengths of approximately $30\,$nG. Meanwhile the use of Faraday rotation measures have resulted in IGMF estimates and limits of $4$-$10\,$nG \citep[e.g][]{Pshirkov16,Osullivan19,Vernstrom19}. More recently, \citet{locatelli21} used the non detection of diffuse emission in filaments connecting two pairs of clusters, using LOFAR HBA observations, to infer upper limits of $\lesssim 0.2 ~\rm \mu G$ for the filament's IGMF.  

Recently \citet{Govoni19} and \citet{Botteon20} were the first to detect diffuse synchrotron emission from intercluster bridges using the LOw Frequency ARray (LOFAR), with peaks at the mJy level. These two, as the only detected thus far with current generation radio telescopes, likely represent the peak of the distribution of filaments in terms of brightness caught in a ``short" ($\leq 1\,$Gyr) phase of high dynamical activity (leading to in situ particle acceleration) before a major merger event \citep{bv20}. Both of these detections were shorter filaments, or intercluster birdges (1-$3\,$Mpc), with \citet{Bonafede20} recently finding a $2\,$Mpc long intracluster bridge within the Coma cluster. However, fainter filaments or those between widely separated clusters with no overlap from cluster emission (which would be the majority of filaments rather than inter or intracluster bridges), in the presence of instrumental noise and source confusion from galaxies, remain below the detectable threshold. Thus different techniques are required to reach below the noise and study the average characteristics of the the population of filaments. 

One method for detecting faint diffuse filamentary emission is to use image stacking. When stacking the faint emission coherently adds, while the noise and uncorrelated emission does not, thus increasing the signal above the noise. This requires a large number of samples, as the noise decreases like $1\over \sqrt{N}$. This requirement could be problematic in terms of filaments as the location of filaments is not well known and even the number of known clusters over the whole sky is only of order thousands. Thus a proxy, or tracer, for clusters with much greater known numbers is needed.

Luminous red galaxies (LRGs) are known to be powerful tracers of large-scale structure. These are massive early-type galaxies that usually reside in, or near, the centres of galaxies clusters or groups \citep{Hoessel80,Schneider83,Postman95}. With the Sloan Digital Sky Survey (SDSS), over a million LRGs have been identified with photometric or spectroscopic redshift. Thus pairs of LRGs that reside near each other in on the sky and in physical space can act as a proxy for physically nearby pairs of clusters, which may be connected by inter-cluster bridges or filaments. The technique of stacking physically co-located pairs of LRGs to study filaments has already lead to detections and measurements of the thermal Sunyaev-Zel'dovich (tSZ) effect in filaments \citep{deGraaff19,Tanimura19}, the mass of filaments from weak lensing \citep{Clampitt16,Epps17}, and dark matter mass-to-light ratios \citep{Yang20}. However, as yet, this technique has not been applied to radio maps in order to look for synchrotron emission from filaments, which is the focus of this work.  

In this work we use multiple radio and X-ray maps to see if a stacked filament detection can be made by stacking pairs of LRGs near each other in physical space, following the method laid out by \citet{deGraaff19} and \citet{Tanimura19}, to examine what such a detection can tell us about the magnetic fields of the intergalactic medium. Section~\ref{sec:data} details the radio and X-ray maps used as well as the LRG sample. In Sec.~\ref{sec:method} we detail the stacking procedure, as well as the procedures for modelling the non-filamentary component and for handling of point sources. Section~\ref{sec:results} reviews the results of the stacking. In Sec.~\ref{sec:discussion} we discuss the results in terms of the possible instrumental causes, systematics and biases, the point source contribution along with possible diffuse shock and dark matter interpretations of the results. Throughout this work, we adopt a $\Lambda$CDM cosmology from \citet{Planck16XIII} with $\Omega_{\rm m} = 0.3075$, $\Omega_{\Lambda} = 0.6910$, and $H_0 = 67.74\,$km s$^{-1}$ Mpc$^{-1}$ for conversion of redshifts into distances and define the spectral index, $\alpha$ such that the observed flux density $I$ at frequency $\nu$ follows the relation $I_{\nu}\propto \nu^{+\alpha}$.

\section{Data}
\label{sec:data}

\subsection{Radio \& X-ray Maps}
\label{sec:raddata}
Ideally for this type of analysis low frequency radio data is desired, as diffuse filamentary emission is expected to be steep spectrum with $\alpha \le -1$ \citep{Vazza15}, with a large sky area also being desirable. With these criteria in mind there are a few options for available data. First is the data from the GaLactic and Extragalactic All-sky Murchison Widefield Array (GLEAM) survey \citep{Wayth15,Hurley-Walker17}. The GLEAM survey covers the entire sky south of Dec +30 at frequencies ranging from $72\,$MHz to $231\,$MHz. The compact array configuration of the Murchison Widefield Array \citep[MWA,][]{Tingay13} during the GLEAM survey provided very good instantaneous {\it uv}-coverage translating to a good sensitivity to large-angular scale emission.  From the GLEAM survey we use three maps: GLEAM Blue with $139 \le \nu \, [{\rm MHz}] \, \le 170$, GLEAM Green with $103 \le \nu \, [{\rm MHz}] \, \le 134$, and GLEAM Red with $72 \le \nu \, [{\rm MHz}] \, \le 103$. 

One limitation of the GLEAM survey is lack of coverage in the northern sky. Thus for additional sky and frequency coverage, and to diversify the data used in hopes of mitigating possible instrument- or processing-dependent effects we also use data from the Owens Valley Radio Observatory Long Wavelength Array \citep[OVRO-LWA,][]{Eastwood18}. The OVRO-LWA data covers the sky north of Dec $-30\,$degrees at frequencies ranging from $36\,$MHz to $73\,$MHz. From this survey we use only the $73\,$MHz map. 

\begin{figure*}
\centering
\includegraphics[scale=0.4]{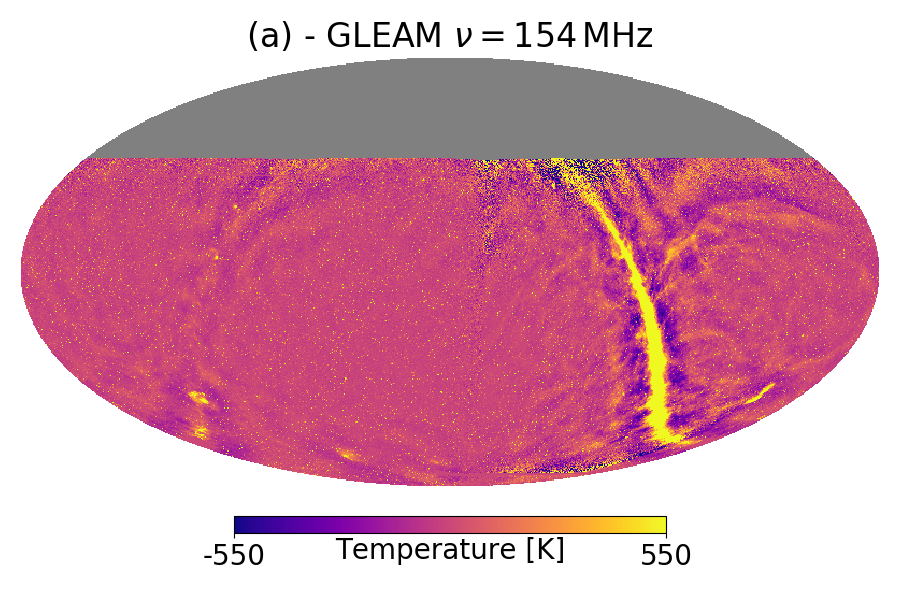}\includegraphics[scale=0.4]{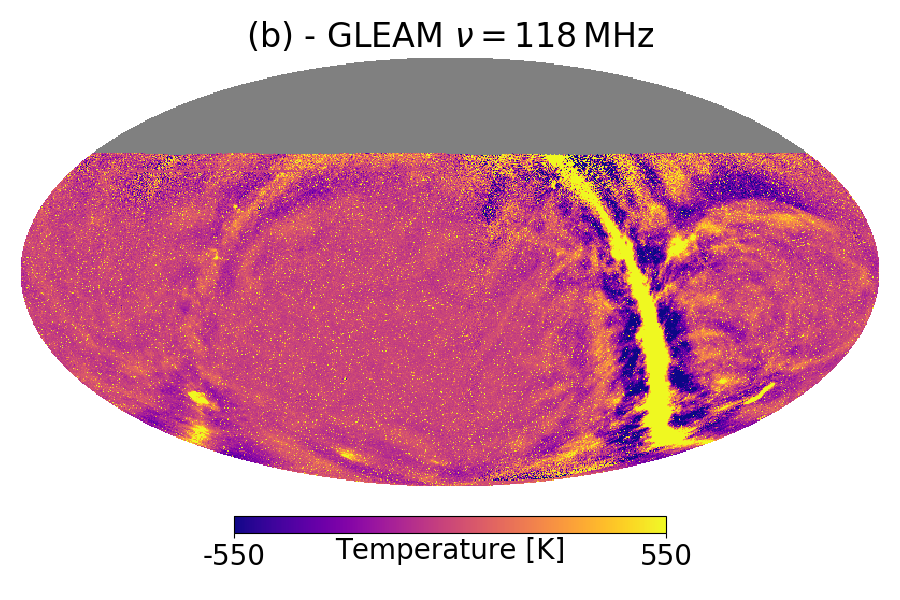}
\includegraphics[scale=0.4]{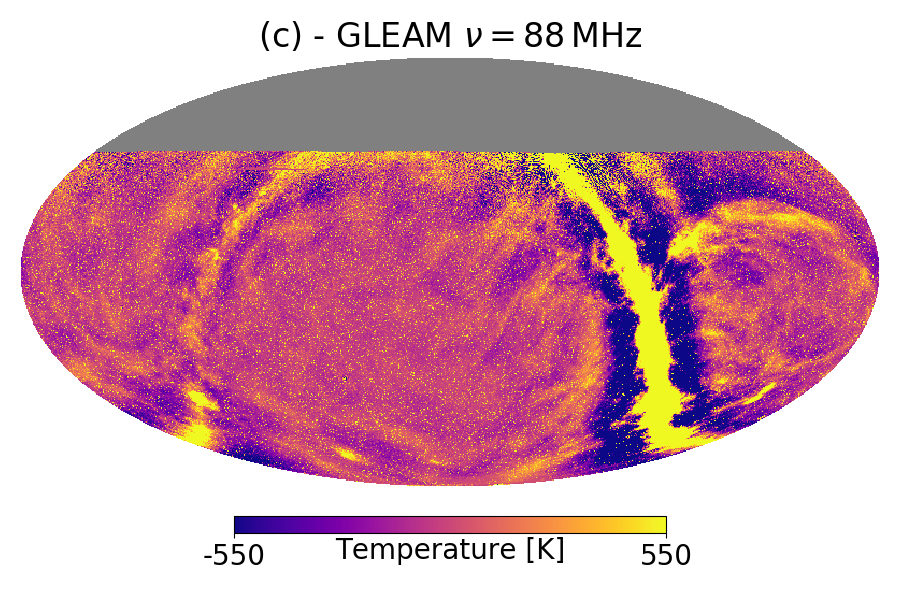}\includegraphics[scale=0.4]{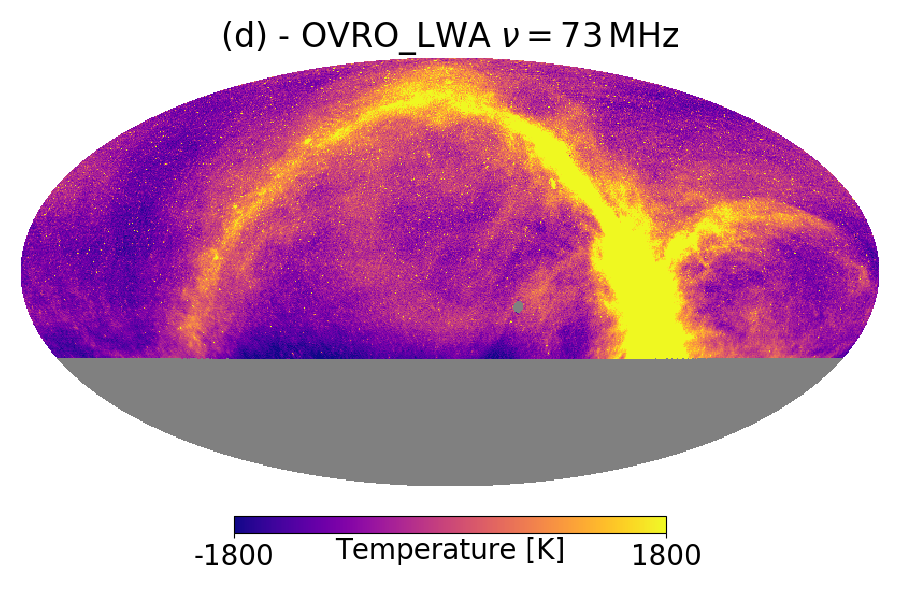}
\includegraphics[scale=0.4]{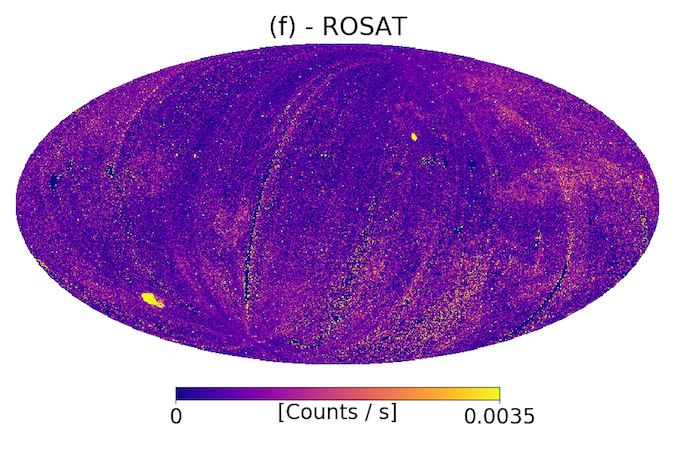}
\caption{The four radio maps  and one X-ray map used in this analysis. From left to right, top to bottom: GLEAM $154\,$MHz, GLEAM $118\,$MHz, GLEAM $88\,$MHz, OVRO LWA $73\,$MHz, and ROSAT, all in equatorial coordinate system where 0,0 is at the centre.}
\label{fig:radmaps}
\end{figure*}

Summary details of the four maps used are listed in Table~\ref{tab:raddat}, with images of each shown in Fig.~\ref{fig:radmaps}. The GLEAM data was mosaicked, converted to brightness temperature, and put onto a Healpix \citep{Gorski05} grid with $N_{\rm side}= 4096$ (correpsonding to a pixel resolution of $0.86\,$arcmin), while the OVRO-LWA data was already available in Healpix format (Galactic coordinates) with $N_{\rm side}= 2048$ (pixel resolution of $1.7\,$arcmin). Healpix maps are well-suited for working with full-sky maps. 

\begin{table}
\centering
 \setlength{\tabcolsep}{3.9pt}
\caption{Details of the radio maps used for this analysis. The beam FHWM and RMS columns are average values as these depend on sky position. The last column gives the approximate largest angular scale (LAS) that the instruments are sensitive to at each frequency.}
\label{tab:raddat}
\begin{tabular}{lrrrrr}
\hline
\multicolumn{1}{c}{Name} & \multicolumn{1}{c}{$\nu$} & \multicolumn{1}{c}{FWHM} & \multicolumn{2}{c}{RMS} & \multicolumn{1}{c}{LAS}  \\
 & \multicolumn{1}{c}{\scriptsize [MHz]} & \multicolumn{1}{c}{\scriptsize [arcmin]} &  \multicolumn{1}{c}{\scriptsize [K]} &  \multicolumn{1}{c}{\scriptsize [mJy beam$^{-1}$] }&  \multicolumn{1}{c}{\scriptsize [Degrees] }\\
 \hline
GLEAM Blue &154 & 2.75 & 20 & 35 & 14 \\
GLEAM Green &118 & 3.6 & 40 &75 & 19 \\
GLEAM Red &88 & 5 & 80 & 143 & 25 \\
OVRO LWA &73 & 16 & 154 & 598 & 47 \\
\hline
\end{tabular}
\end{table}

The X-ray data used was from the The ROentgen SATellit {\it ROSAT} All Sky Survey \citep[RASS,][]{Truemper92,Truemper93}. We stacked on the total energy band map (0.1 - $2.4\,$kev), as well as the individual hard energy band map (0.4 - $2.4\,$kev) and the soft energy band (0.1 - $0.4\,$kev). The ROSAT resolution is approximately $1.8\,$arcmin. The Healpix maps have Nside=2048. The X-ray maps are converted to surface brightness units, or counts s$^{-1}$ arcmin$^{-2}$, using the Healpix pixel area.

\subsection{Catalog}
\label{sec:optdata}

As a tracer of large-scale structure we use the Luminous Red Galaxy (LRG) catalogue from the Sloan Digital Sky Survey (SDSS) Data Release 7 \citep{Lopes07} (with a total number of LRGs in the catalogue  $\sim 1\, 400 \, 000$, all with photometric redshifts). Even though the presence of an LRG does not guarantee a cluster (or group), and it is not a guarantee that the LRG is at the centre of a cluster, with enough samples the contribution to the stack from those that are not in clusters or are largely offset from cluster centres should average out, or possibly decrease any detected signal (in which case the true signal from only actual cluster pairs could be higher than what is detected here). 

From all of the computed pairs with a co-moving separation of $\Delta R \le 15\,$Mpc, we select only those with angular separations of $ 20 \le \Delta \theta \, [{\rm arcmin}] \le 180$. The lower limit is set to be slightly larger than the largest beam size of the four radio maps. The upper limit is set to keep from having too large of a range of $\Delta \theta$s interpolated onto the same grid. From this we have a total of $390\, 808$ LRG pairs. 

\begin{figure}
\centering
\includegraphics[scale=0.37]{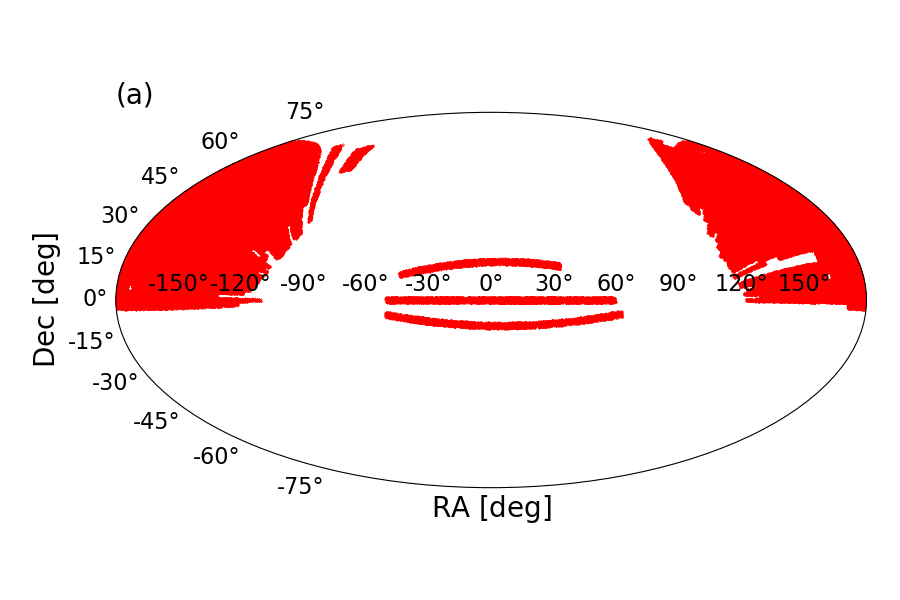}
\includegraphics[scale=0.37]{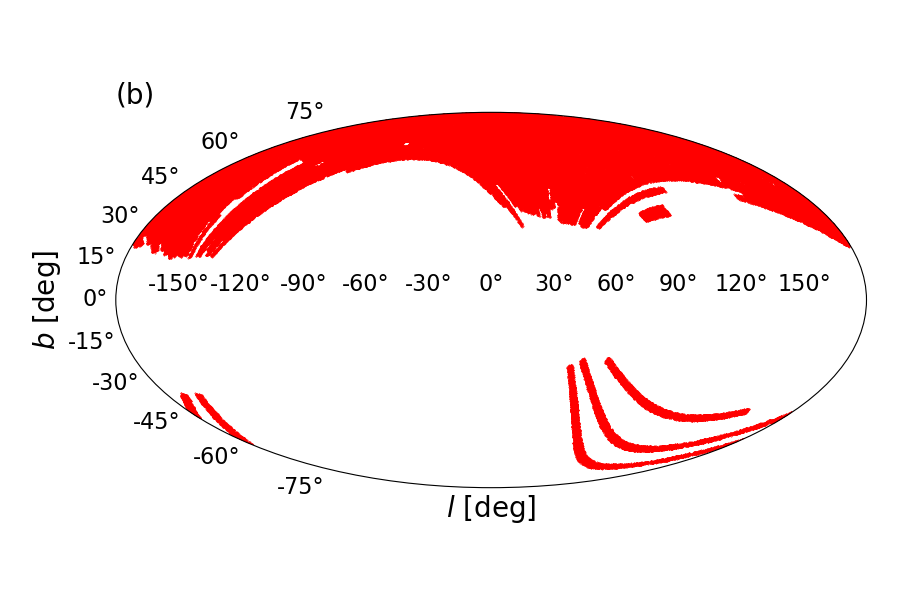}
\caption{Sky positions of the 390,808 LRG pairs. Panel (a) shows in celestial coordinates and panel (b) is Galactic coordinates. }
\label{fig:lrgpos}
\end{figure}

\begin{figure*}
\centering
\includegraphics[scale=0.27]{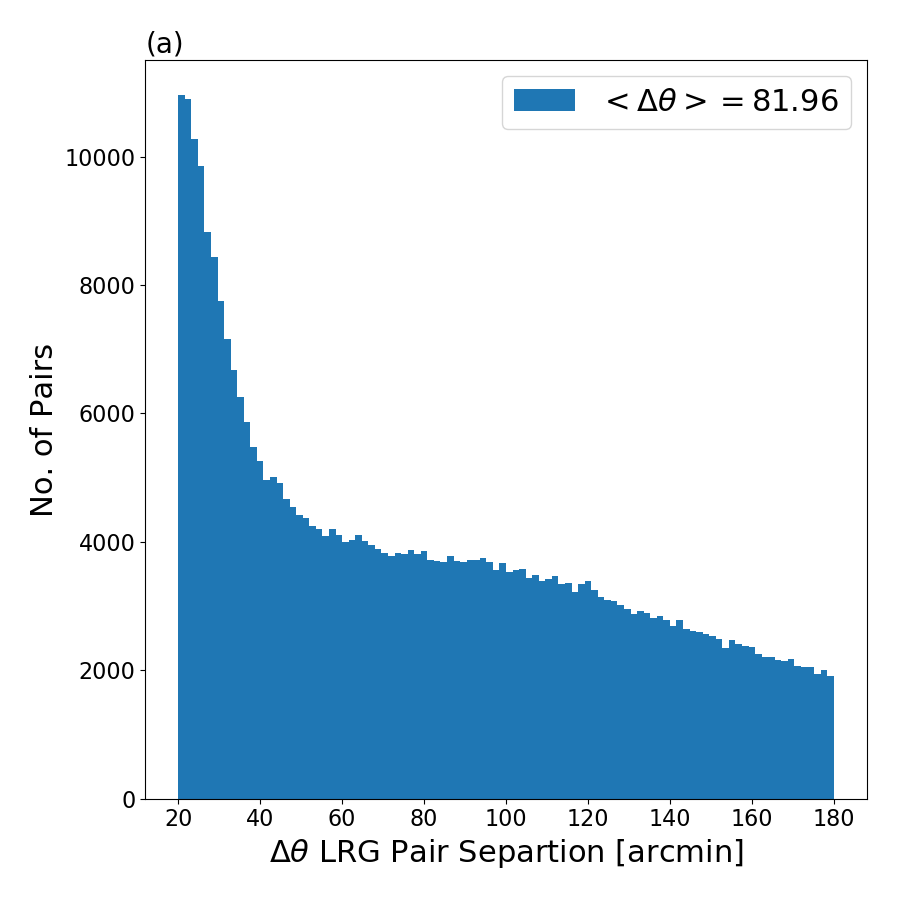}\includegraphics[scale=0.27]{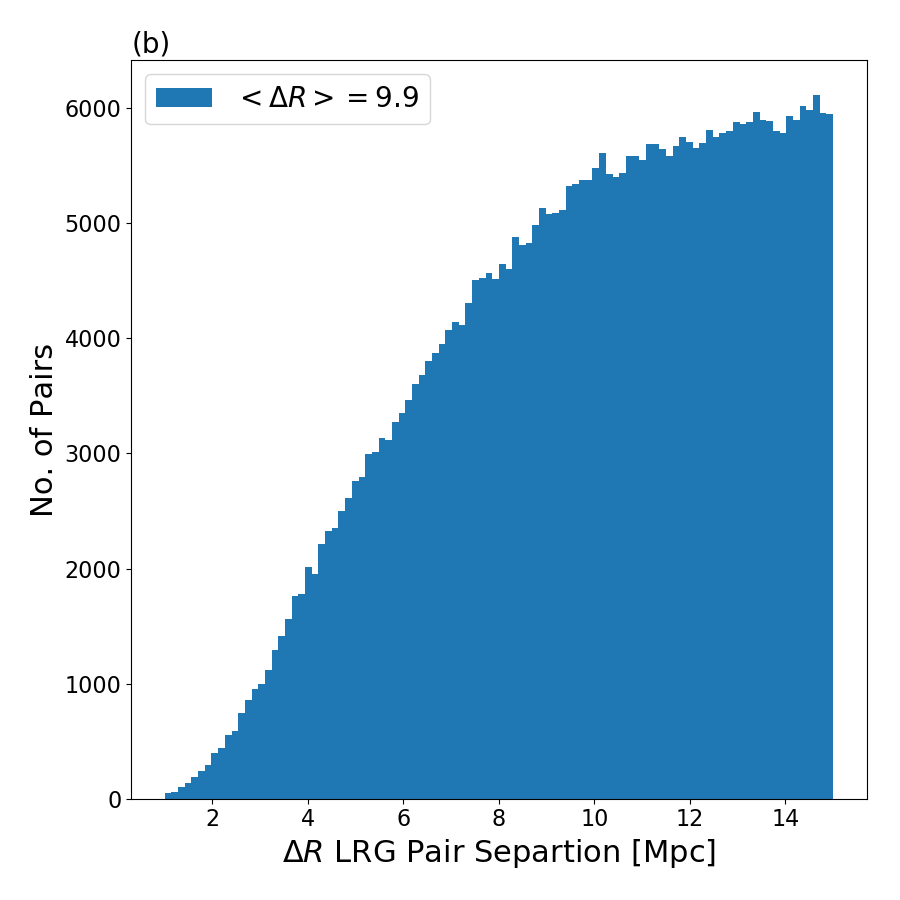}\includegraphics[scale=0.27]{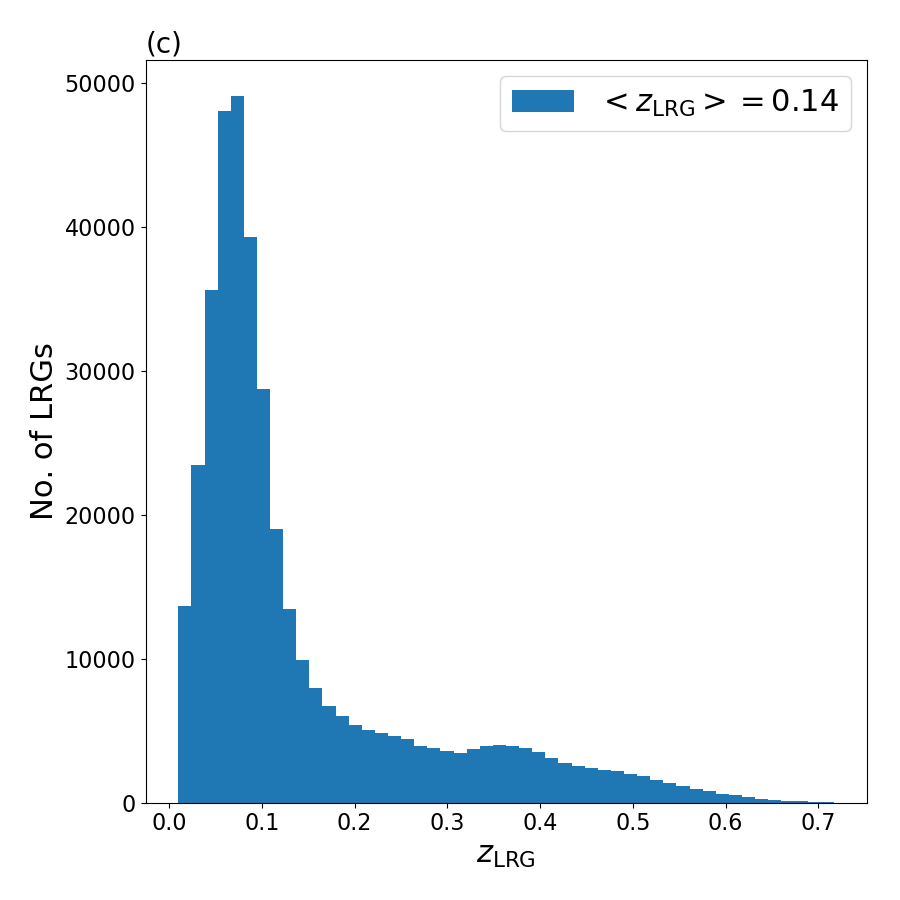}
\caption{Distributions for the 390808 physically nearby pairs in the stacking sample. Left panel shows the separation of the LRGs in angular distance, the middle panel shows the distance in co-moving Mpc, and the right panel shows the average redshift of the LRG pair.}
\label{fig:realhists}
\end{figure*}

The sky distribution of the LRG pairs is shown in Fig.~\ref{fig:lrgpos}. The separation distributions, both angular and physical, as well as the redshift distribution, are shown in Fig.~\ref{fig:realhists}. The average angular separation of the pairs is $< \Delta \theta > =82\,$arcmin, with the average co-moving physical separation $< \Delta R > \simeq 10\,$Mpc. The average redshift of the pairs is $<z>=0.14 \pm 0.01$, with a maximum redshift of $z_{\rm max} = 0.716$.  

\section{Method}
\label{sec:method}

\subsection{Stacking}
\label{sec:stack}

\begin{figure}
\centering
\includegraphics[scale=0.35]{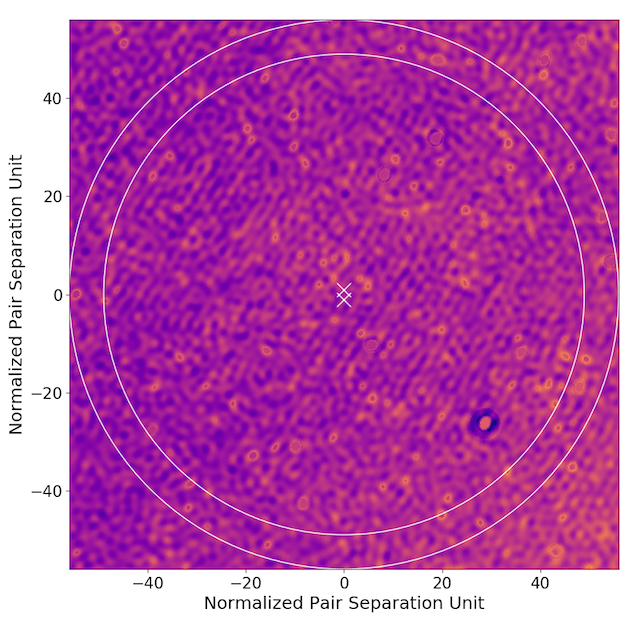}
\includegraphics[scale=0.35]{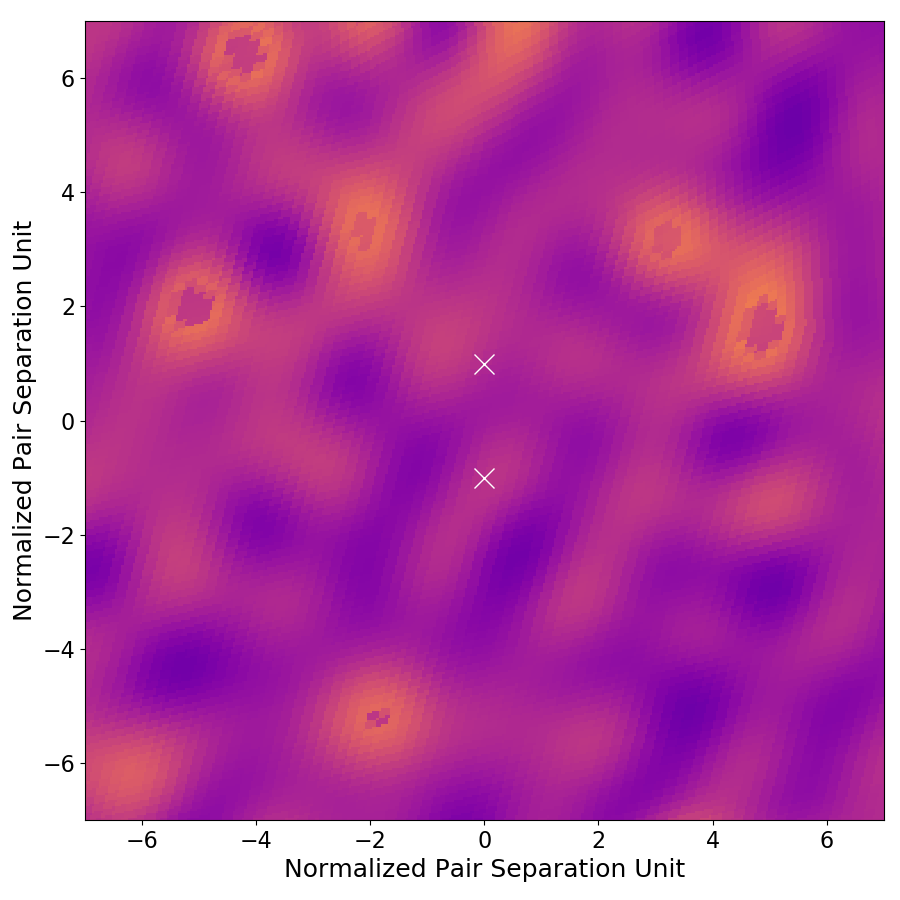}
\caption{One cutout from the stack of the OVRO-LWA data showing positions of the LRGS and region used for background estimation. The bottom panel shows the regular grid around the two LRG positions (white x's) while the top panel shows a larger region around the LRGs with the region between the two white circles used to estimate the average the background value.  }
\label{fig:tback}
\end{figure}

We follow the method for stacking laid out in similar filament stacking works \citep[e.g.][]{Clampitt16,deGraaff19,Tanimura19}. The angular separation between LRG pairs in our catalogue ranges between $20$ and $180\,$arcmin. For each pair, we follow \citet{Clampitt16} and make a two-dimensional cutout, or stamp, around the two LRGs. This cutout is transformed onto a normalised 2D image coordinate system, (X, Y), with one LRG placed at (0, $-1$) and the other placed at (0,$+1$). This requires both a scaling of the pixel sizes as well as a rotation to align the two LRGs along the vertical axis. This scale factor and rotation angle will be different for each individual cutout. The transformation from sky coordinates to a normalised grid is applied to the each cutout map. Then the average background signal in the cutout is estimated. The mean signal in the annular region $9 < r < 10$, where $(r^2 = X^2 + Y^2)$, is subtracted as an estimate of the local background (an example of this background region is shown in one case in Fig.~\ref{fig:tback}.) This is done to ensure the mean value of each map in the stack is approximately zero. After the transformation and background subtraction, each cutout is added to the previous cutouts. A weight image is also made for each cutout. The weight consists of ones and zeros; one where a pixel is valid (or not NaN or inf values) and zero if the pixel is invalid). The weight maps are also summed. The final stacked images are then made by dividing the image sums by the weight sums. Through the regridding and rescaling process the surface brightness of the emission is conserved. 

We also examine a possible different weighting scheme. In previous similar filament stacking studies, the weights for each image cutout in the stack are assigned as simply 1 or 0, for either a valid pixel or not valid. We decided to test if using a variance weighting scheme would impact the results, using the measured variance in the region of each LRG pair. 

We computed this variance weighted stack for each dataset and found no significant change (within uncertainties) of the measured values, the measured signal to noise, or the background noise of the final images. This is confirmation that the noise values around the LRG pairs do not vary significantly. 

\subsection{Point Sources}
\label{sec:psources}
Point sources do not stack like Gaussian noise. Confusion noise from faint point sources will stack like instrumental noise, in that the noise from unresolved or blended faint sources will go down as $1\over \sqrt{N}$ in the stack. However, bright sources, those above the instrumental and confusion noise values, will not easily average out and can skew stack values. In this work, we are not interested in bright radio sources and did not want them to obstruct the stacking; thus we needed to find some manner of subtracting them or mitigating their contribution. 

There is a GLEAM survey catalog. However, creating a model map from this catalog and subtracting out the sources could leave areas of potential over or under subtraction if the model is not entirely accurate. These inaccurate subtractions could bias the results. Also, the GLEAM catalogue only covers a portion of the Northern sky, leaving no model for more northern sources in the OVRO-LWA map. This would also require accurate spectral indices for all the sources.

Another option that we considered was simply masking sources, or pixels, above some threshold. This works to some degree. However, the local background around sources is quite variable over the maps and therefore setting a hard limit would still leave sources that are ``bright", i.e. above the noise limit, but potentially in a negative or lower background region. This could also have the effect of possibly masking pixels from diffuse emission that happen to be brighter or located within a bright background region. 

The best approach we found to remove unwanted point sources was to use wavelets. An invertible isotropic undecimated wavelet transform (UWT) on the sphere \citep{Starck94,Starck06} can be used to transform the Healpix map into a set of wavelet scales where the sum of the wavelet scales reproduces the original image. The transform decomposes the sum of the emission into wavelet scales highlighting emission features at that corresponding scale size, thus allowing us to easily find point source emission and subtract it.

To perform the wavelet transform we used the Interactive Sparse Astronomical Data Analysis Packages (ISAP), specifically the \textsc{mrs$\_$uwttrans} program.\footnote{\url{http://www.cosmostat.org/software/isap}} Once broken down into the different wavelet scales, the smallest scales were used to isolate point sources. The rms of the scale map was found and then all pixels $> 5\sigma$ were put into a model. Once the three smallest scales were searched the total model was subtracted from the original radio map (or the sum of the scales). An example of one zoomed in region of the OVRO-LWA map and the GLEAM Green map, before and after, this process is shown in Fig.~\ref{fig:pssub}. For additional details on this process, see Appendix~\ref{sec:Wavelet Model}.

As a test, the stacking procedure was performed on maps with nothing done regarding the point sources, a blanking mask based on pixel brightness set to $5\sigma$, and the wavelet subtracted maps. The final output was not significantly different, with the main difference being the addition of bright sources where no subtraction or masking was applied.

\begin{figure*}
\centering
\includegraphics[scale=0.33]{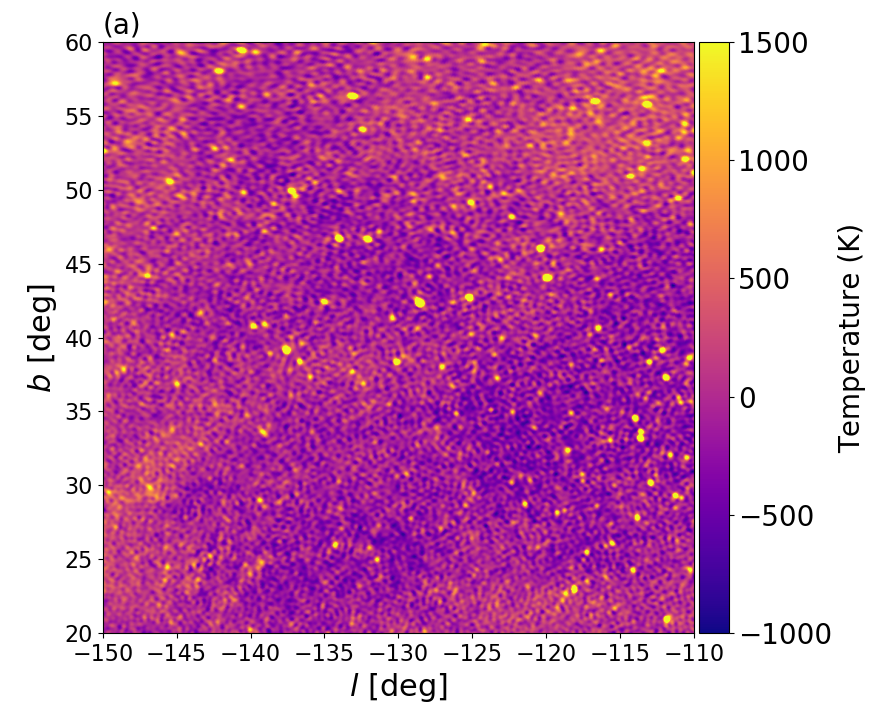}\includegraphics[scale=0.33]{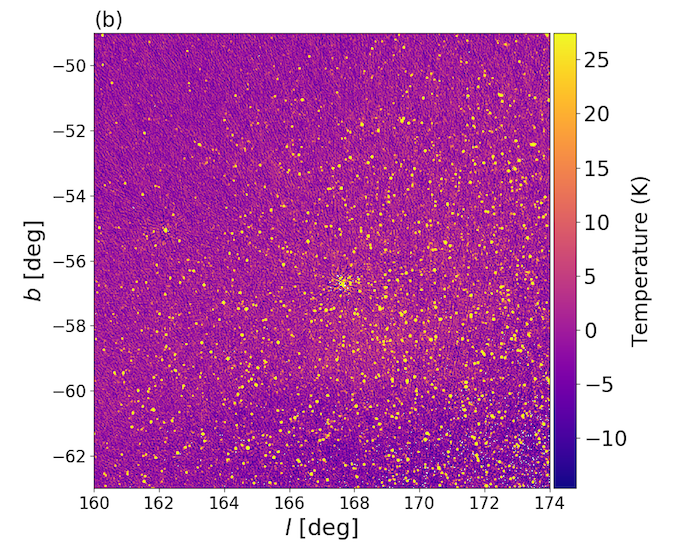}
\includegraphics[scale=0.33]{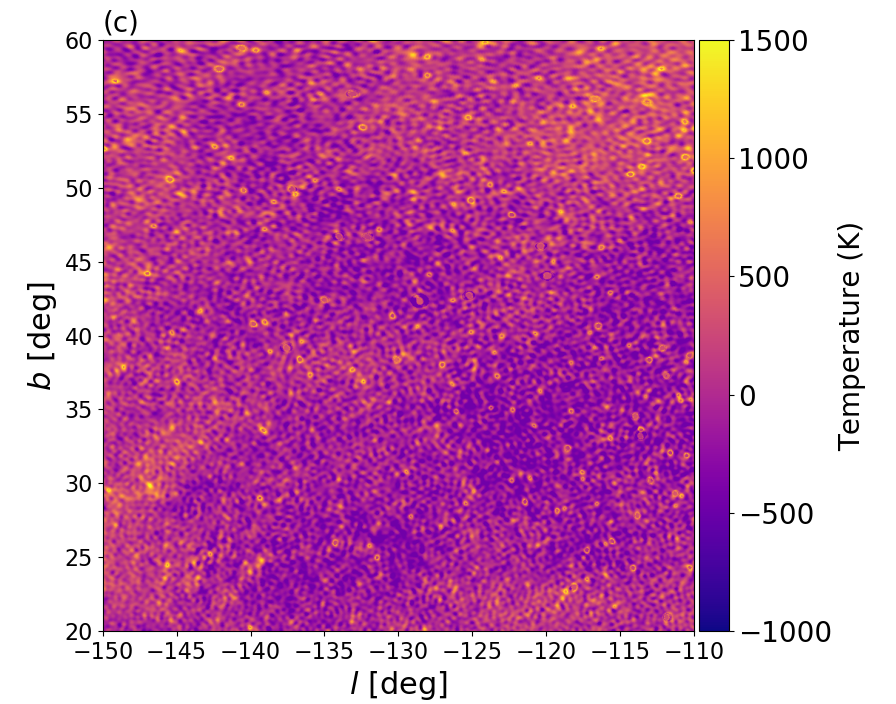}\includegraphics[scale=0.33]{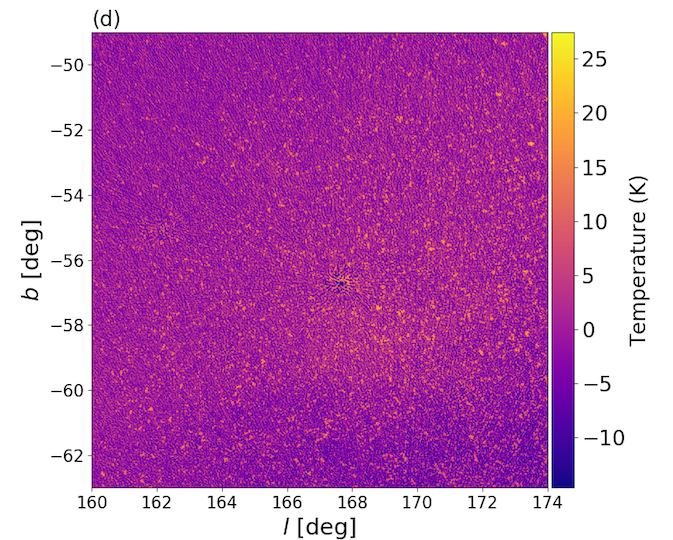}
\caption{Sections of radio images before and after wavelet point source filtering.Panels (a) and (b) are before filtering and panels (c) and (d) are after, with both top and bottom panels having the same scaling. The left panels, (a) and (c), are sections of Owens Valley $73\,$MHz map, while the right panels, (b) and (d), are the GLEAM Green, or $118\,$MHz map.}
\label{fig:pssub}
\end{figure*}

\subsection{Subtraction of Halo Contribution}
\label{sec:subtract}
\begin{figure*}
\centering
\includegraphics[scale=0.27]{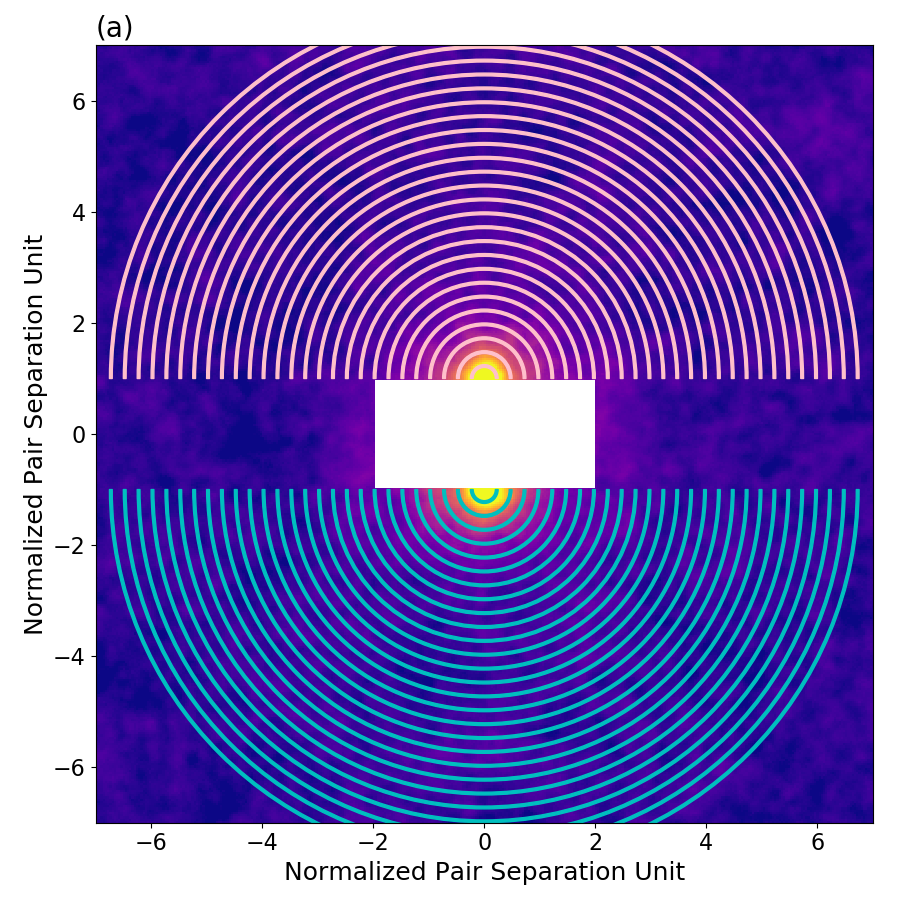}\includegraphics[scale=0.27]{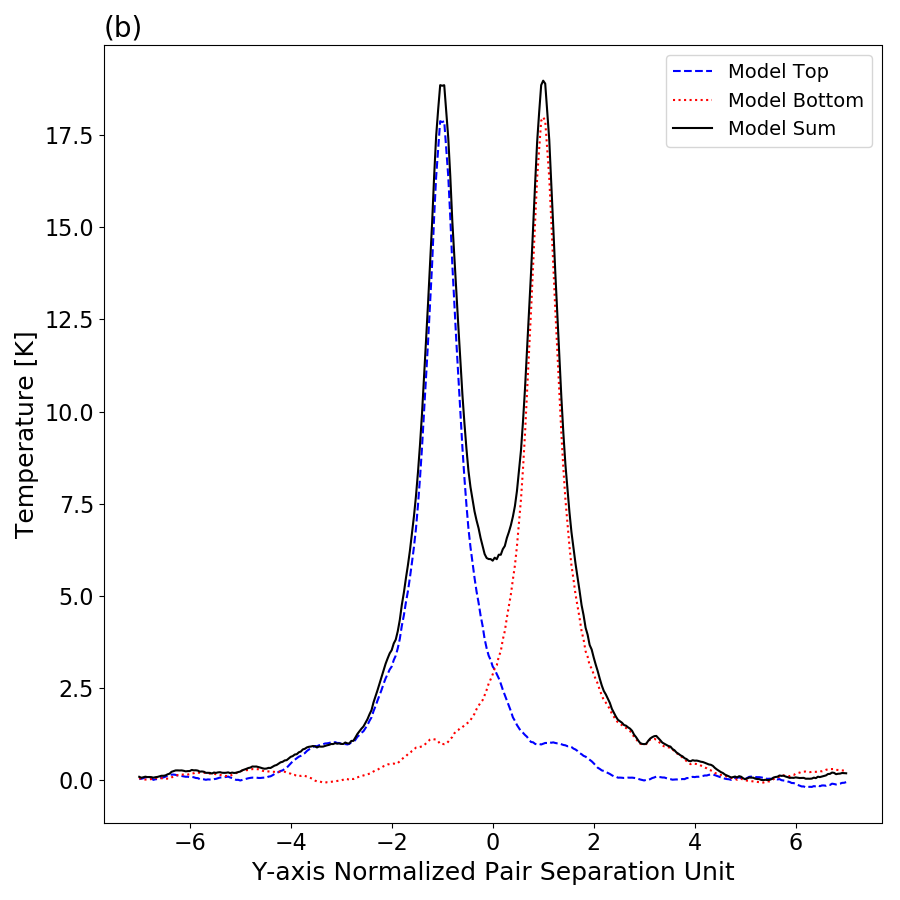}\includegraphics[scale=0.27]{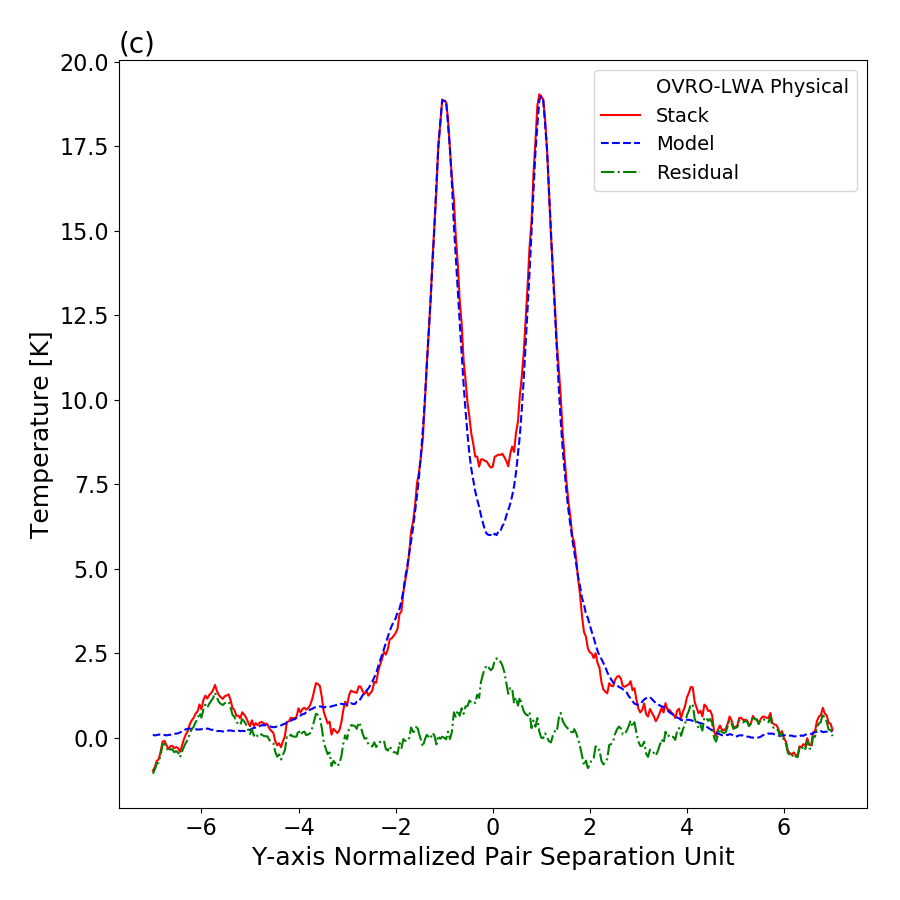}

\caption{Demonstrations of the halo modelling and subtraction. The left panel shows the stack image from the OVRO-LWA data, where the white section in the middle is masked out. The colored semi-circles show that the model is found by taking the radial average around each of the central points individually to come up with two radially averaged models. The sum of the top and bottom models is subtracted from the stack, The middle panel shows the 1D profile through the center of the image at $x=0$ for the top (red dotted line) and bottom (blue dashed line) models and the sum of the two (black solid line).  The right panel shows the same 1D profile for the stacked image (red solid line), the model (blue dashed line), and the residual (green dot-dashed line). }
\label{fig:model}
\end{figure*}

We assume that there is a halo, or non-filament, contribution around each LRG. This may include the central LRG radio or X-ray counterpart as well as any additional diffuse halo emission or excess cluster emission due to faint point sources. We make the assumption that this non-filament emission is radially symmetric around each LRG. We take the radial average around each LRG, excluding the region between the pair to make a model for each LRG's non-filament emission. The model for each LRG is added together to create the combined model. This model is then subtracted from the stack image to obtain the excess, or filament emission. Figure~\ref{fig:model} shows an example of the modelling process and results. 

\subsection{Resampling \& Null Tests}
\label{sec:null}
\begin{figure*}
\centering
\includegraphics[scale=0.27]{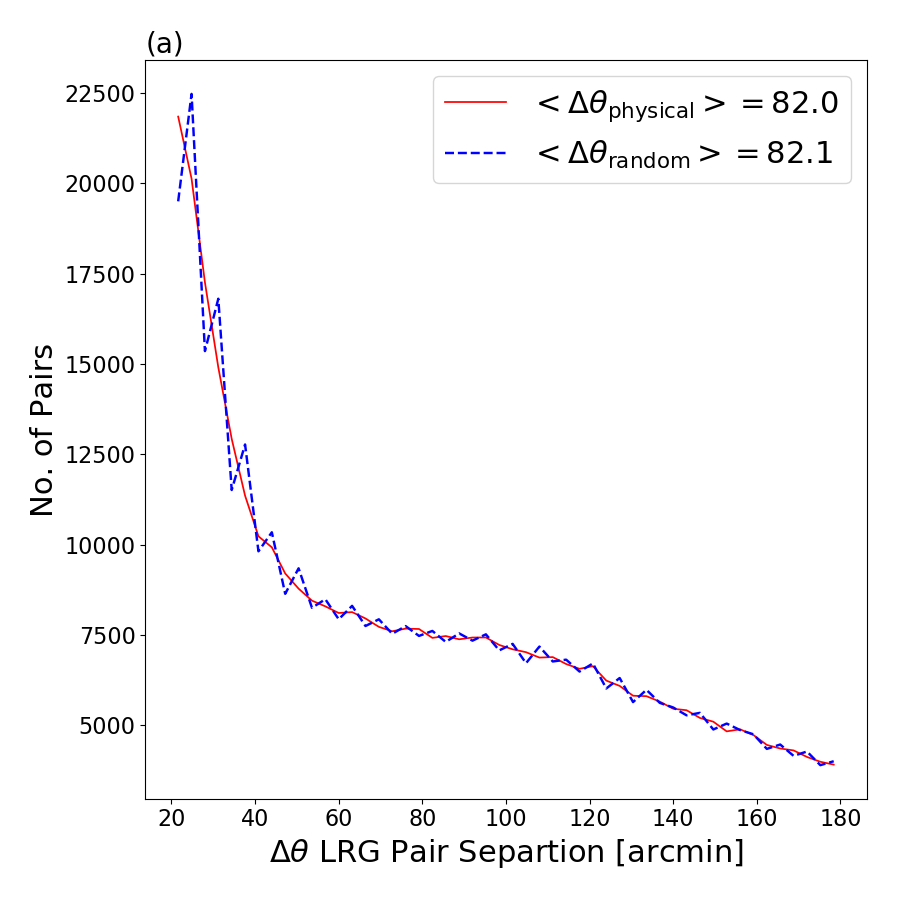}\includegraphics[scale=0.27]{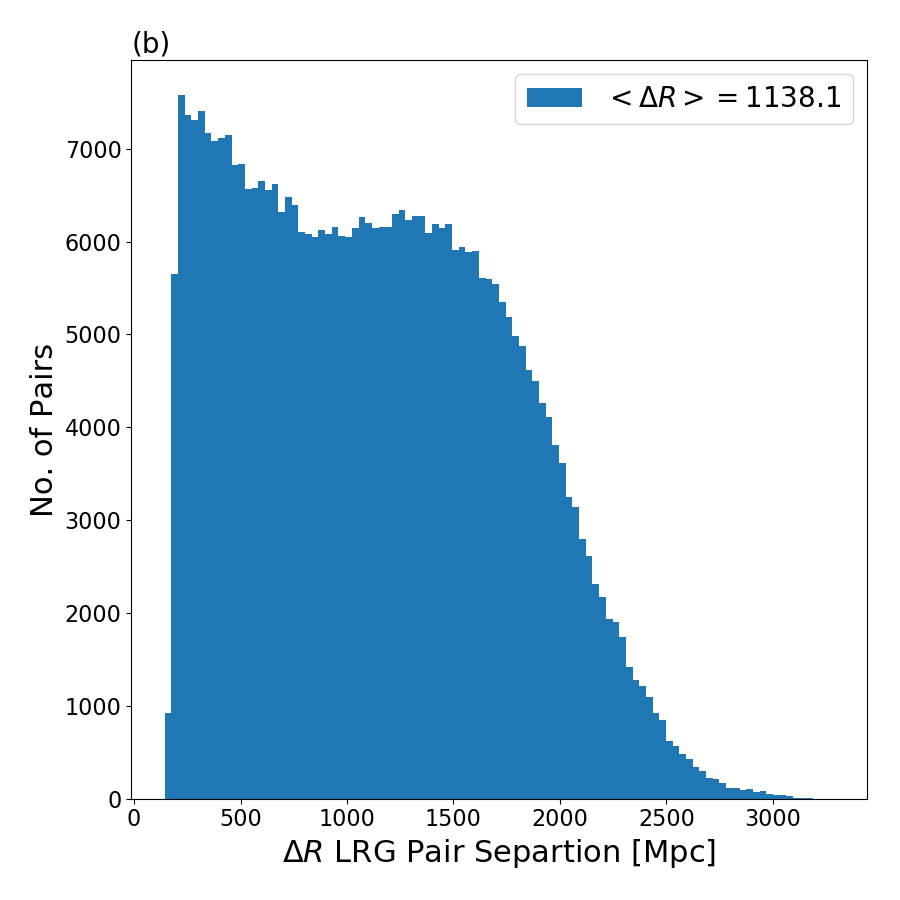}\includegraphics[scale=0.27]{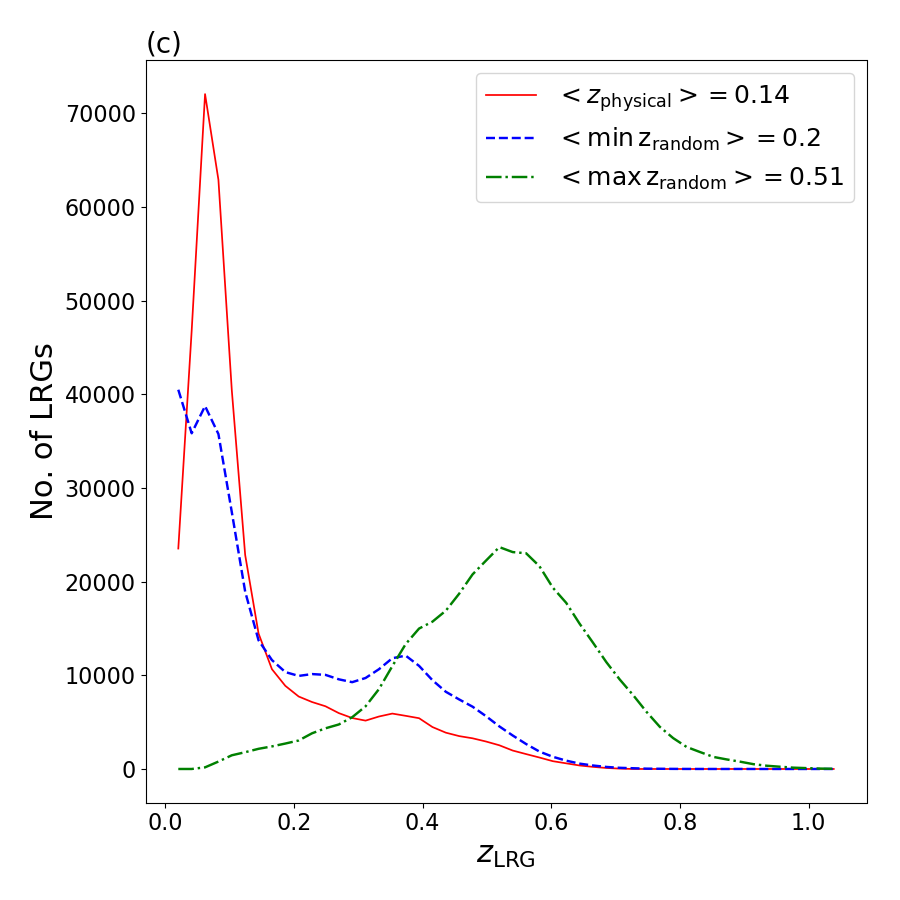}
\caption{Distributions for the 390,808 pairs used in the physically nearby LRG sample and the randomly associated null test sample. Left panel shows the separation of the LRGs in angular distance (red solid line is physically nearby pairs and blue dashed line is randomly associated pairs), the middle panel shows the distance in co-moving Mpc for randomly associated pairs, and the right panel shows the average redshift of the LRG physically nearby pairs (red solid line) and the minimum redshift (blue dashed line) and maximum redshift (green dot-dashed line) for the randomly associated pairs.}
\label{fig:randhists}
\end{figure*}

In order to see whether there is any significant emission in the stack between clusters, or LRGs, that are physically nearby one another, the most obvious test is to compare to similar samples that are only randomly associated with each other, or LRG pairs that are ``nearby" each other on the sky (small angular separation) but with large physical separation. Using the LRG catalogue, we create unique samples of pairs of the same $N_{\rm LRG}$ (or $390\, 808$ LRG pairs) with the same angular separation distribution as the physical pairs, but with the added condition that the $\Delta z \ge 0.05$. This ensures that the sample, drawn from the same original sample of sources in the same sky region, has the same number of pairs with the same angular separation distribution but the physical separation, in co-moving Mpc, is much larger.  The LRGs in these pairs will have different redshifts from each other (due to the criterion that they be far apart in physical space), but in order to try and closely match the sample of physical LRG pairs we added the condition that the redshift distribution of at least one LRG in the pair closely match that of the physically related LRG pair's redshift distribution. The distributions for one such randomly selected sample is shown, in comparison to the physical pairs, in Fig.~\ref{fig:randhists}.

For each of the radio maps and X-ray maps, 500 null stacks are generated (or 500 stacks made from $390\, 808$ randomly selected LRG pairs). For each null stack map, a model and residual map were also created. These maps are referred to as the ``control" sample in the results.  

As a second null test, we stacked on $N_{\rm LRG}$ blank fields by using the centre position of the physically related pairs and applying a randomly generated shift of up to 3 degrees, with random direction. These cutouts are then not centred on any sources in particular but contain a similar patch of sky as the physically related pairs and should therefore capture any systematics that pertain to the local areas around the physically related LRG pairs. For these random ``blank" stacks 500 sets of $N_{\rm LRG}$ were also created for each radio and X-ray dataset. 

It is also important to test the in-sample variance of the physically related LRG pairs. To do this we performed a bootstrap test wherein we randomly resampled (with replacement) $75\,$per cent of the $N_{\rm LRG}$ from the physically related pairs and recomputed the stack, the model, and the residual. We repeated this resampling 500 times for each of the radio and X-ray maps.  The bootstrap test was also done using $50\,$per cent and $90\,$per cent of the pairs, with similar results.

To test the average sky background level from these maps using this method, we selected $N_{\rm LRG}$  blank fields, or fields selected at random from the same region of the sky as the LRG sample (but not centred on any known sources) and performed the same stacking procedure. These blank field stacks were repeated 500 times for each map as well. While the control group mentioned above tests the scenario of stacking un-physically related LRG pairs from the same LRG population, this tests the more for any sky variance issues in the region or potential instrumental effects from the data.

\section{Results}
\label{sec:results}

\begin{figure*}
\hspace*{-.3in}
\includegraphics[scale=0.67]{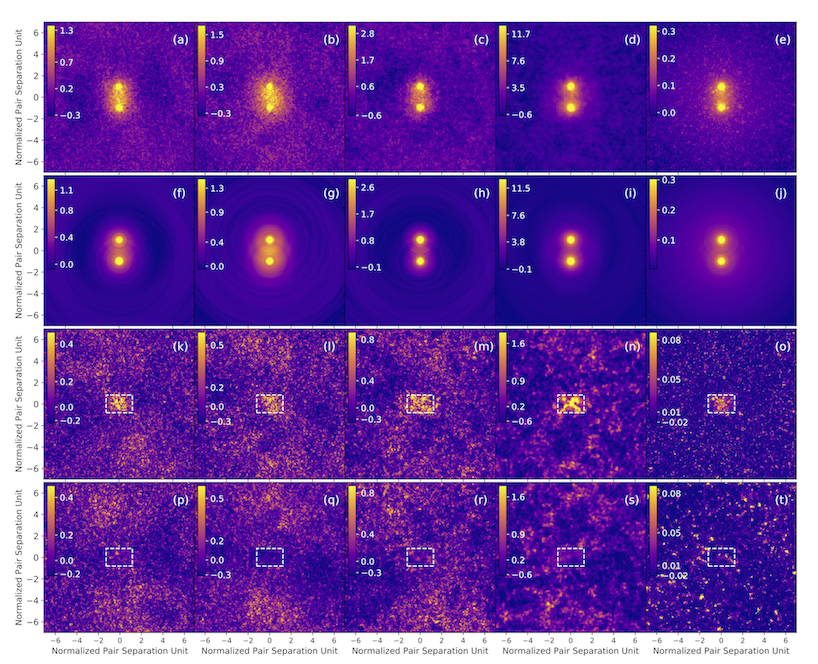}
\caption{ Stacking results for physically nearby pairs. Top row is the final stack images, second row is the model images, third row are the residual images, and the bottom row is the residual images from the control sample (with the stack and models from the control sample not significantly visually different from the physically related sample). From left to right the columns are GLEAM 154 MHz, GLEAM 118 MHz, GLEAM 88 MHz, OVRO-LWA 73 MHz, and ROSAT combined band 0.1 -2.4 keV. The inset colour bars all have units of temperature in K, except the ROSAT maps which are in (counts per second per arcmin$^{2}$) $\times 10000$. In the bottom row, the white box outlines the region in which the average filament signal is obtained. For the bottom two rows of residuals the colormap is shown with a power-of-two scaling, rather than linear, to better accentuate the features. }
\label{fig:realim}
\end{figure*}

The results from stacking, along with the models and residuals, for all four radio maps and the combined X-ray map can be seen in Fig.~\ref{fig:realim}. Corresponding residuals from the averages of the 500 null stacks are shown in the bottom row (the control or null tests went through the same process of stacking and modelling as the physical pairs stacks, only the residuals are shown in Fig.~\ref{fig:realim}). 

The white boxes (bottom two rows of Fig.~\ref{fig:realim}) highlight the region between the two LRG positions where any excess or filament signal should be present. This region shows clear positive signals in the residual images of the physically related LRG pairs in all maps, whereas there is no significant excess signal for the control sample. The measured surface brightnesses from these regions are listed in Table~\ref{tab:restab}. To test the significance of the average signal in this region, we applied a 2D moving average across the maps the size of the box region. Comparing the average values in the box region against the rms from the averages outside of the central region, we find the average in the box to be significant for all the physical pair maps at the $\geq \, 5\sigma$ level, except for the blue GLEAM map and the soft-band ROSAT (0.1 - 0.4 keV) map which have significance values of $3.1\sigma$ and $3.6\sigma$, respectively.

\begin{figure*}
\centering
\includegraphics[scale=0.31]{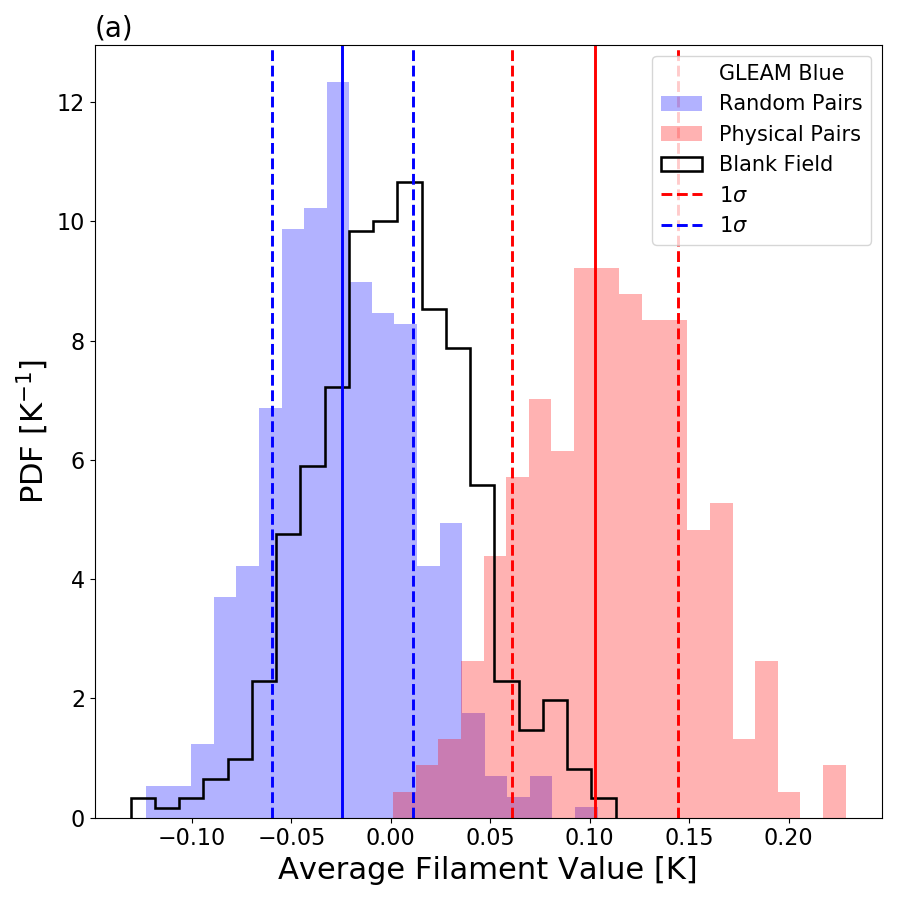}\includegraphics[scale=0.31]{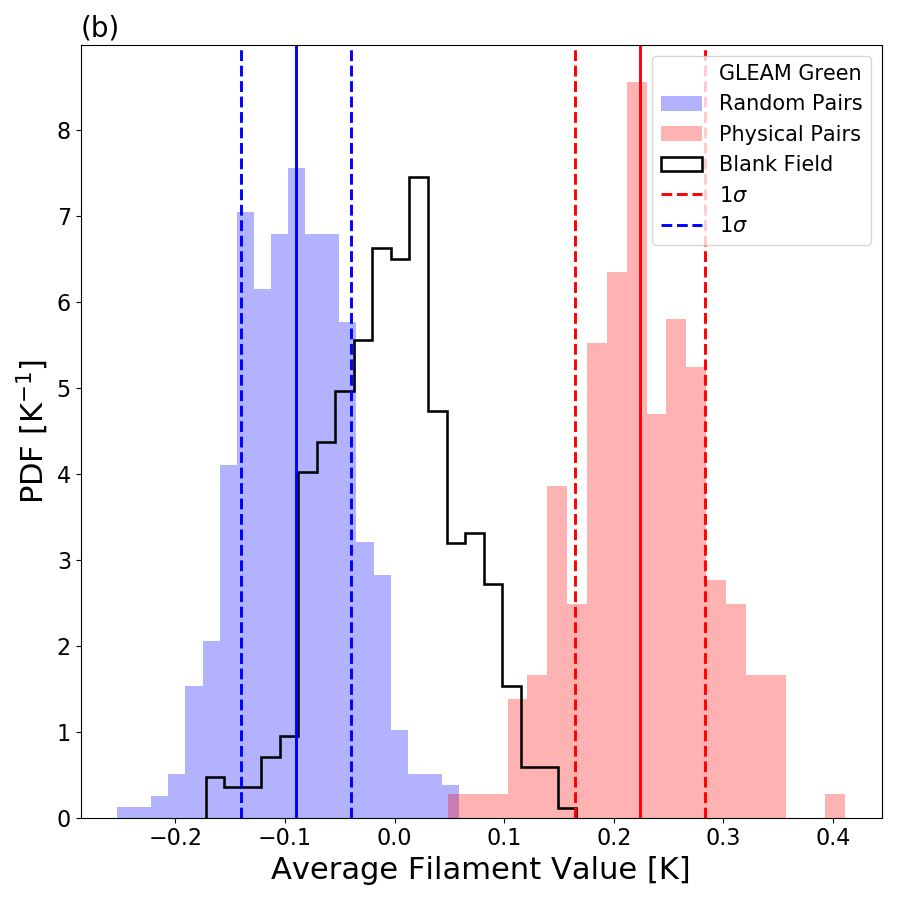}
\includegraphics[scale=0.31]{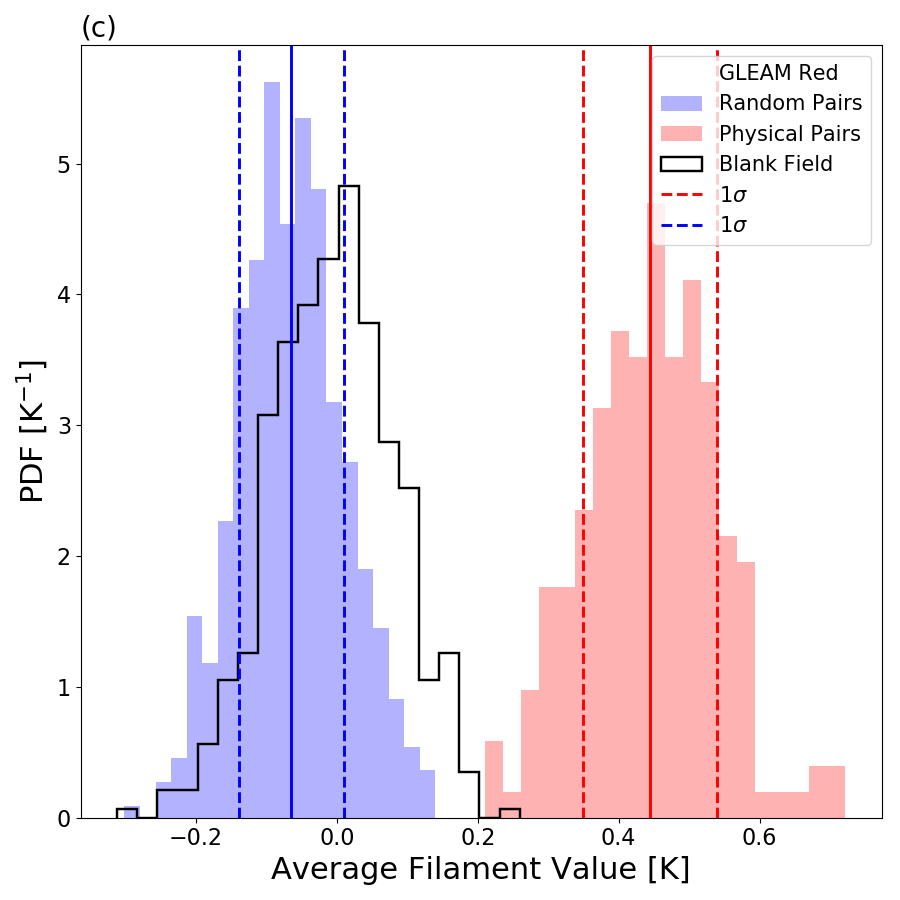}\includegraphics[scale=0.31]{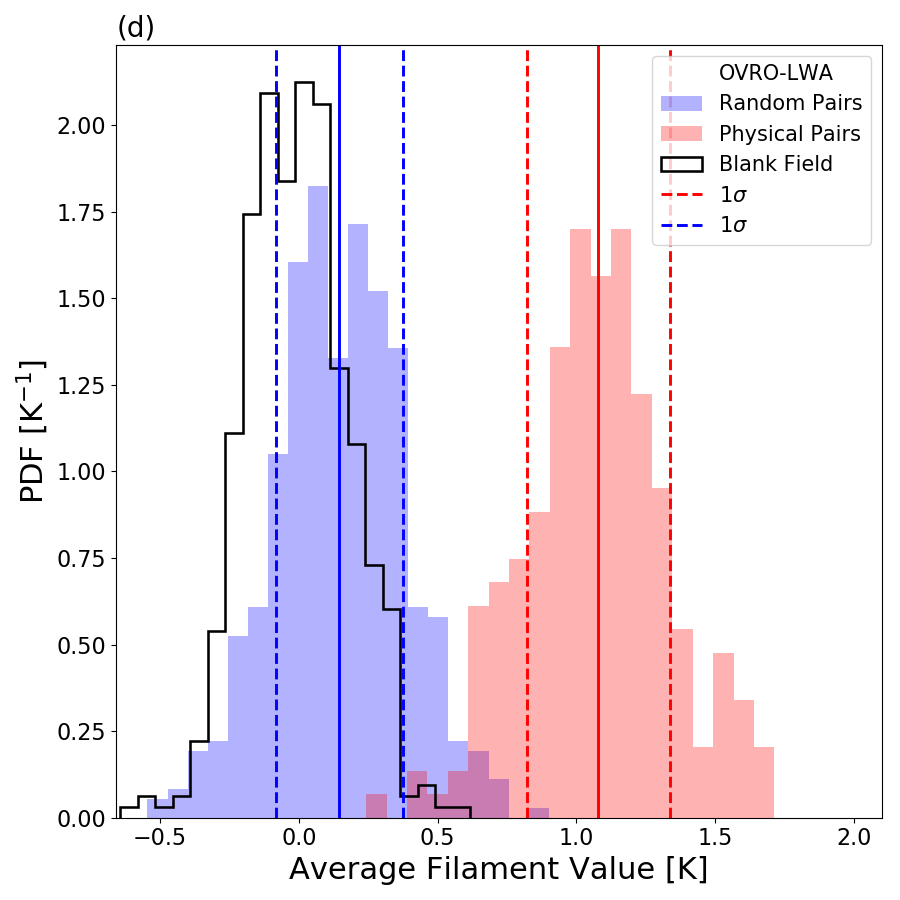}
\includegraphics[scale=0.31]{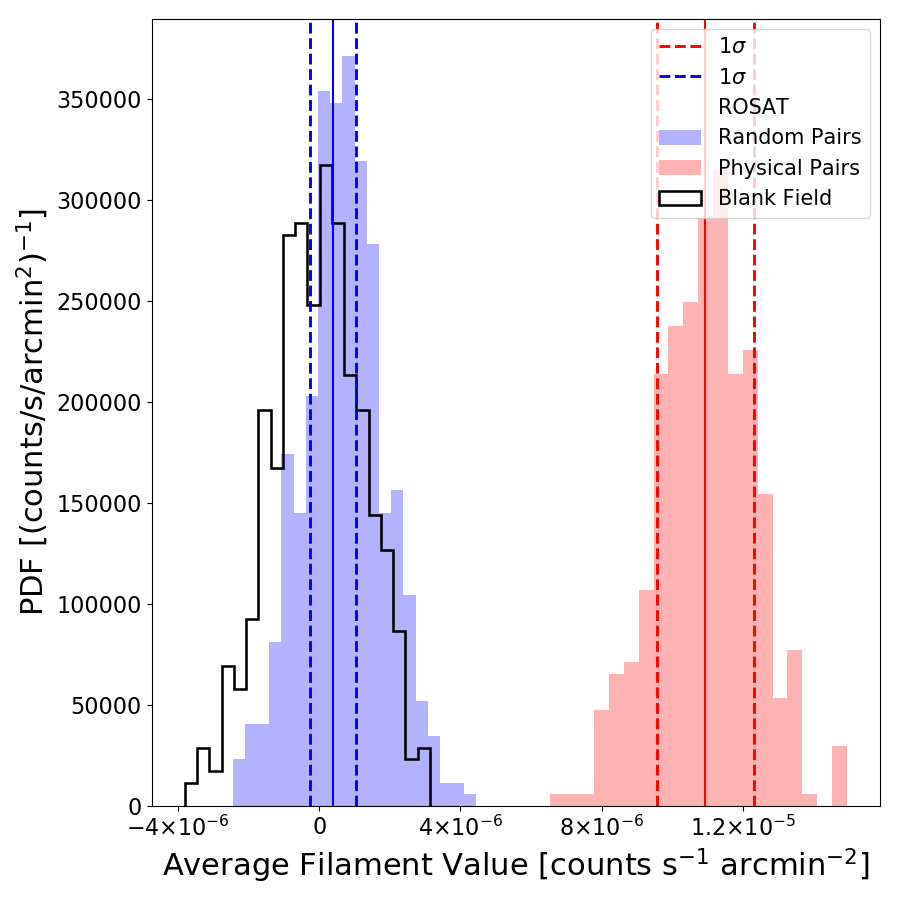}
\caption{Distributions of average filament values from resampling tests. The blue distributions are from the control tests of stacking on randomly associated LRG pairs. The red distributions are from randomly resampling portions of the physically nearby LRG pairs. The black lined distribution is from stacking on blank fields, or fields not centred on any LRG pairs. The vertical solid lines show the mean of the distributions while the vertical dashed lines show the $1\sigma$ regions.}
\label{fig:nulldist}
\end{figure*}

We show in Fig.~\ref{fig:nulldist} the distributions of average signal within that central region from all of the control maps (both the random control maps and blank field maps), as well as all of the bootstrap maps from resampling the physical pairs.. From this it is clear to see that the distributions are significantly different. Two-sample Kolmogorov-Smirnov (KS) test is performed on all of the maps with $p$-values $<< 0.0001$, showing that these are definitely not the same distributions. The width of the distributions, or the $1\sigma$ uncertainties are well matched to the expected noise level of the stacked maps, where the noise in the stacked map is expected to be $\sigma_{\rm stack}=\sigma/\sqrt{N_{\rm LRG}}$. Such expected $\sigma_{\rm stack}$ values are listed in Table~\ref{tab:restab}. 

\begin{figure}
\centering
\includegraphics[scale=0.33]{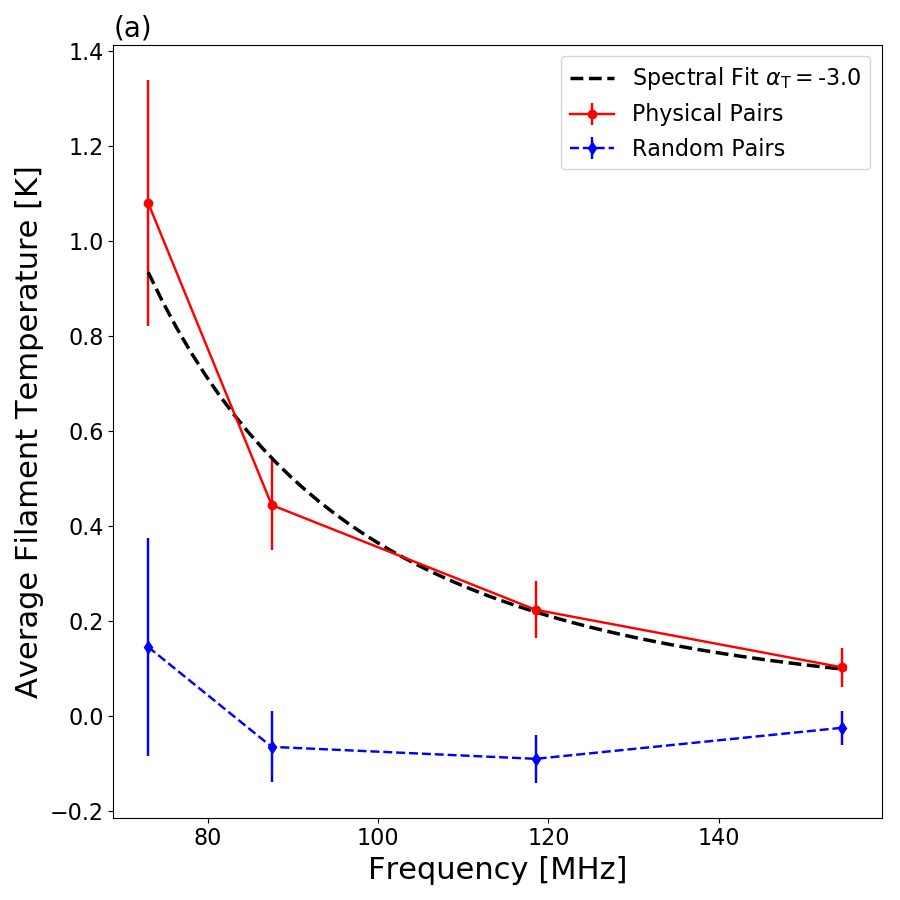}
\includegraphics[scale=0.33]{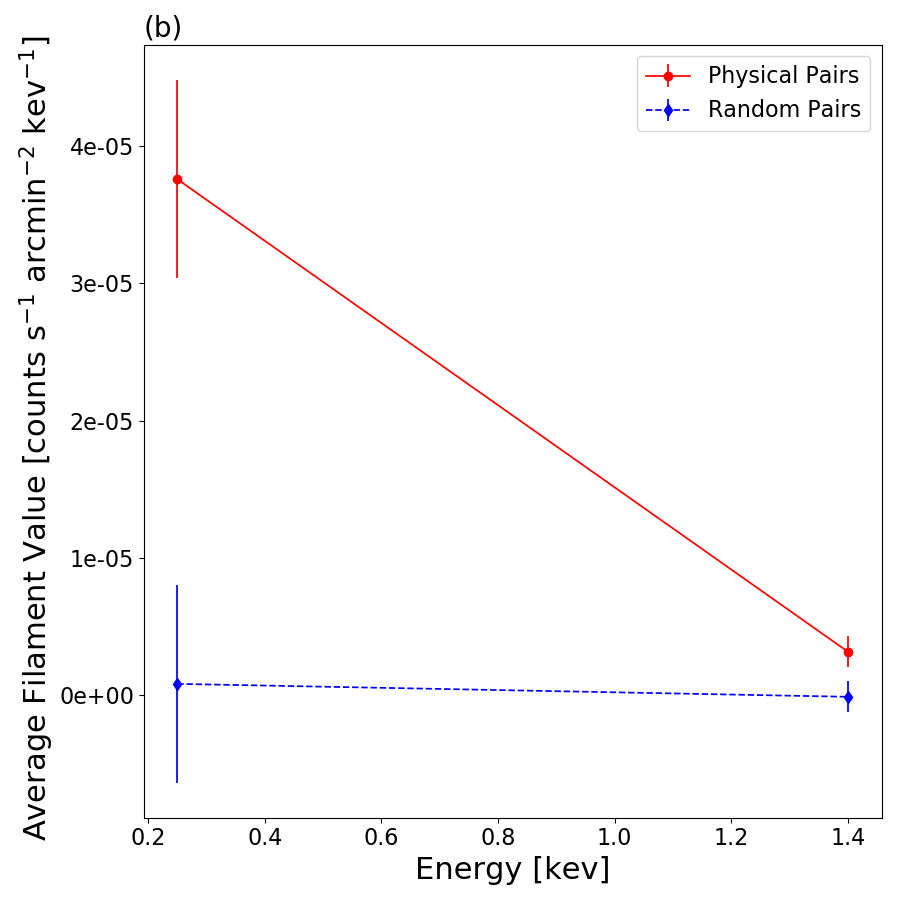}
\caption{Average filament value as a function of frequency for the physically nearby pairs and the average of the randomly associated pairs. The top panel shows the four radio frequencies while the bottom panels shows the two X-ray frequencies. }
\label{fig:fvalsvfreq}
\end{figure}

The physically related pairs show a positive average signal, that increases with decreasing frequency (see Fig.~\ref{fig:fvalsvfreq} and Table~\ref{tab:restab}). Fitting a power-law model to the radio signal of the physical pairs yields a temperature spectral index $\gamma =-3.0 \pm 0.1$, which translates to a spectral index in flux density of $\alpha= \gamma +2 =-1.0 \pm 0.1$ (with $S_{\nu} \propto \nu ^{\alpha}$), which is consistent with observations of strong radio cluster shocks \citep{Vanweeren19}.

We additionally tested subsamples of the physical pairs. We divided the sample into bins of redshift, physical separation of the pairs and angular separation, and repeated the stacking. Subdividing the sample decreases the the number of pairs in the stack, thus increasing the noise. In each case of binning the residual filamentary signal was still seen, however, the signal to noise was too low to look for any trends. 

We investigate the impact of applying a surface brightness correction to account for dimming with redshift. The intensity scales with redshift by a factor of $(1+z)^{\alpha-3}$, where this takes into account both the $k$-correction and cosmological dimming of surface brightness. 

We apply the correction factor to the whole cutout of an LRG pair, meaning any underlying signal but also the noise in that particular cutout is increased by that factor. This means for high-$z$ pairs the noise would be increased by a factor of $\sim 8$, assuming $\alpha =-1.0$. To counter this, we use a weighted sum in this case, with the weighting being inverse variance weighting. However, in this case any $z$-corrections applied will be squared in the variance, i.e. a $z$-correction factor of 8 corresponds to an increase of 64 in the variance, thereby down-weighting the higher-$z$ pairs.

Using two different assumptions for the typical spectral index, $\alpha=-1.0$ and $\alpha=-1.5$, we computed the stacking, using the inverse variance weighting. In both cases the detected signal only changes by $\leq 10\,$per cent across all of the maps, with the resulting spectral index changing by $\sim 15 \,$per cent. Thus, considering the uncertainty in the appropriate value of $\alpha$ and that no significant change is found, we adopt the reported values with no corrections applied.

The power-law index, $\kappa$, between the two ROSAT bands (in counts s$^{-1}$ arcmin$^{-2}$ keV$^{-1}$) yields $\kappa = -0.5 \pm 0.07$. Assuming a power-law model and an average Galactic neutral hydrogen column density of $2\times10^{20}\,$cm$^{-2}$ we obtain an X-ray surface brightness flux for the combined band measurement of $\sim 2.5 \times 10^{-16}\,$ erg s$^{-1}$ cm$^{-2}$ arcmin$^{-2}$. We also used an Astrophysical Plasma Emission Code (APEC) model \citep{Smith01} with 0.2 solar metallicity abundance and $0.9\,$keV Temperature which yields $1.1 \times 10^{-16}\,$erg s$^{-1}$ cm$^{-2}$ arcmin$^{-2}$. \footnote{ The X-ray count rate conversion was performed using the PIMMS webtool, \url{https://heasarc.gsfc.nasa.gov/cgi-bin/Tools/w3pimms/w3pimms.pl}}

\section{Discussion}
\label{sec:discussion}

\subsection{Systematics \& Biases}
\label{sec:bias}

Before discussing the physical interpretation of the detected excess we first examine some areas of possible systematic or instrumental effects that could cause or bias the results. One important thing to consider, particularly with radio interferometers, is the point spread function (PSF), sometimes known as the ``dirty" beam. This dirty beam is what the map is convolved with below some ``clean" threshold. The dirty beam will have positive and negative sidelobes and the size and shape of the beam are dependent on the antenna array configuration, frequency, observing declination, length of observation and parameters defined during the imaging process. Above the clean threshold the brighter sources are convolved with a clean beam, usually a two-dimensional Gaussian fitted to the main lobe of the dirty beam. 
Assuming there are radio sources associated with the clusters or LRGs that would be convolved with these dirty beams, a possible cause of the detected signal could be from the beam patterns coherently adding together.

\begin{figure}
\centering
\includegraphics[scale=0.33]{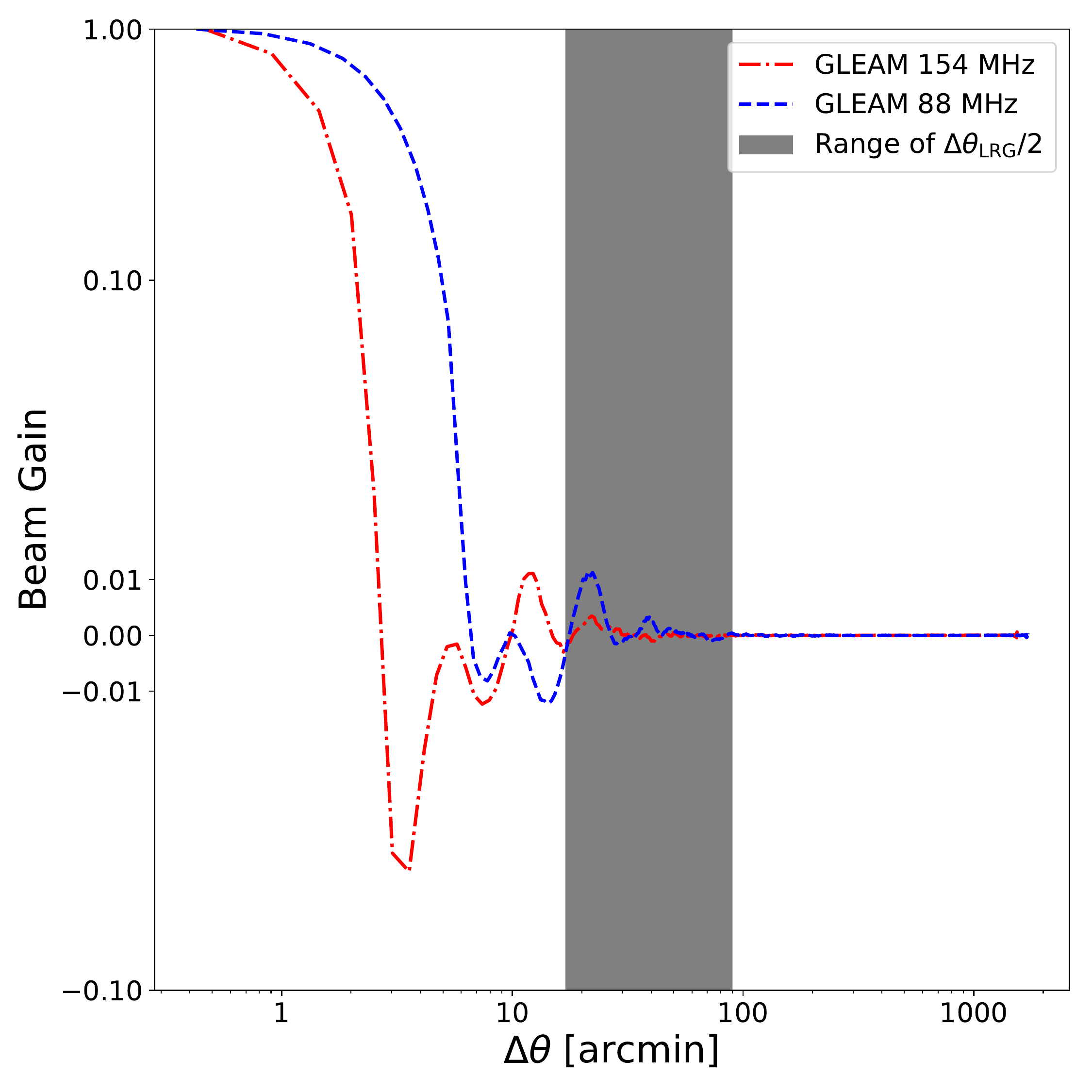}
\caption{Beam profiles of the GLEAM 154$\,$MHz beam (red dot-dashed line) and GLEAM $88\,$MHz (blue dashed line). The grey shaded region shows the approximate range of distances that would fall in the filamentary region between the two LRGs in a pair.  }
\label{fig:beams}
\end{figure}

The peak sidelobes of the dirty beams have maximum amplitudes of $2\,$per cent to $\sim 10\,$per cent depending on the frequency, instrument, and declination. These main peaks tend to lie close in the central lobe, being only separated by $1$ to $\sim 10\,$arcmin. The beam profiles for the highest and lowest GLEAM frequency bands are shown in Fig.~\ref{fig:beams} \citep[see figure 3 of][ for the OVRO-LWA beams]{Eastwood18}. In the region that is half-way between the LRG pairs (or $\Delta \theta_{\rm LRG}/2$ ) the maximum sidelobe values are 0.003 for $154\,$MHz and $0.01$ for $88\,$MHz, with average values in that region of $0.0003\,$arcmin$^{-1}$ and $0.002\,$arcmin$^{-1}$, respectively. Whereas, the surface brightnesses measured in the region between filaments for these two bands corresponds to $\sim 1\,$per cent of the peaks in the $154\,$MHz stack image and $\sim 5\,$per cent of the peaks in the $88\,$MHz stack image. Thus even if the beams added perfectly coherently for the two LRGs (or double the single beam contribution) it would not equal the detected excess (and that is assuming the beam sidelobes would add perfectly coherently which would not be the case). Additionally, PSF sidelobes would not explain why the signal is consistently seen across all the frequency bands, from multiple different instruments, with completely different beam functions and seen to have a spectral index close to that expected for a physical signal. It also would not explain the signal detected in the X-ray or the tSZ maps, which again have completely different PSFs.

Another consideration is the fact that not all of the LRGs may be at the centres of groups or clusters. It is estimated that as few as $75\,$per cent of LRGs are central to a cluster for halo masses of $10^{14.5}\,$M$_{\odot}$ \citep{Hoshino15}. Rather than restrict our original sample size by imposing a mass threshold, which would increase the likelihood of the LRG being central to large mass halo but decrease our sample size, we tested the effect of non-centrality by running repeated stacks with the physical pairs sample but applying an artificial shift to the positions. For each source in each pair we randomly shifted the central stack position from the LRG position by $\pm 2\,$Mpc (assuming an average cluster diameter of $1$-$3\,$Mpc). This test was run multiple times using the OVRO-LWA and the GLEAM 118$\,$MHz maps. There was no significant change to the resulting excess signal. Thus we can conclude that moderate shifts, or off-centre positions of the LRGs, do not play a significant role. Additionally, the fact that some LRGs may not reside in massive haloes or clusters means those pairs would not be contributing to the stack and would decrease the average, making any detection more of a lower limit. 

In the modelling and subtraction step described above in Sec~\ref{sec:subtract}, we assume isotropy, or a radial symmetry, in the regions outside of the area between the two LRGs. For individual clusters this is not a valid assumption both considering the distribution of galaxies inside the cluster, the flux distribution of galaxies inside the cluster, and non-galactic emission (e.g. radio relics). However, once the individual images are stacked the approximate isotropy can occur relatively quickly. 

\begin{figure}
\centering
\includegraphics[scale=0.53]{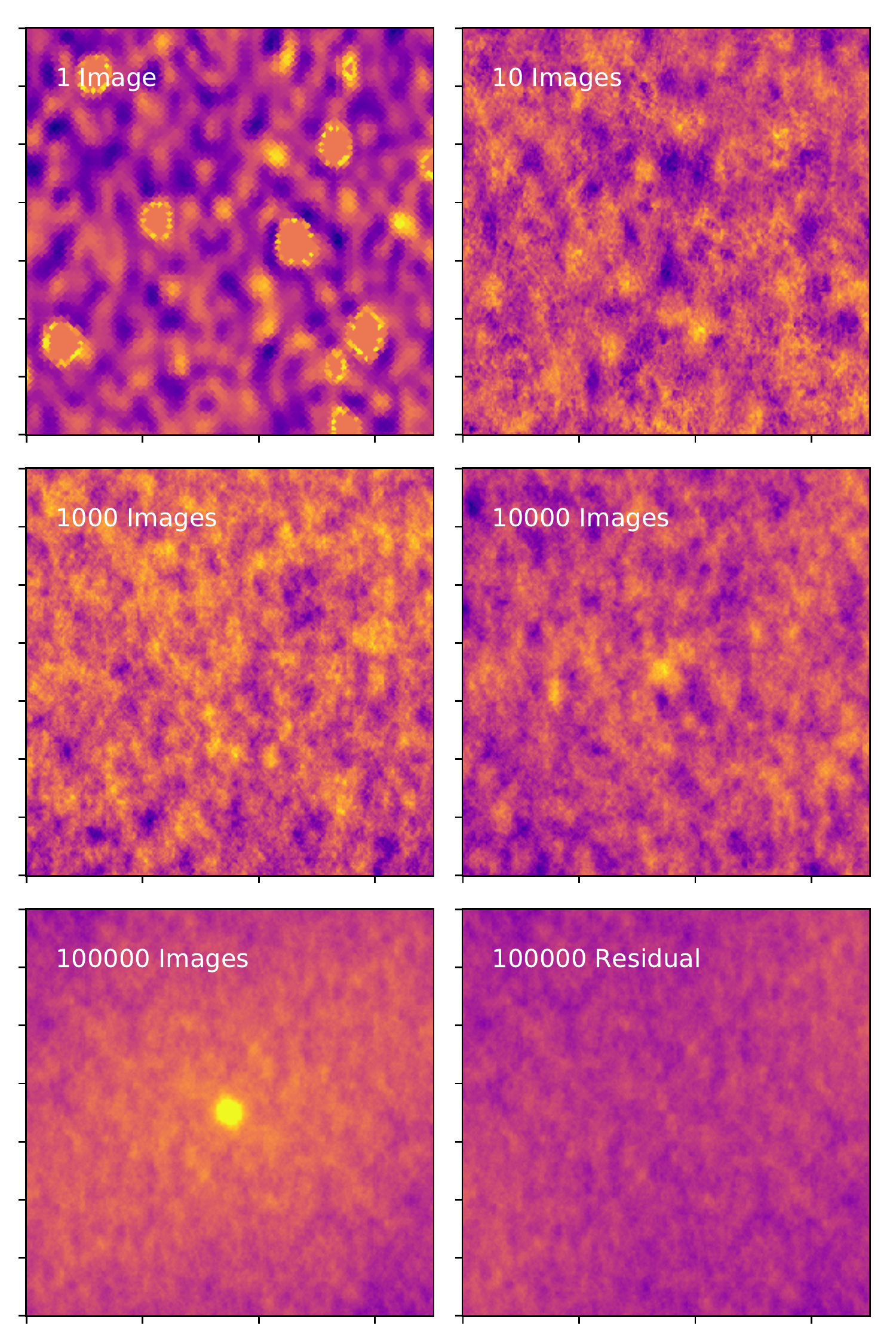}
\caption{Stack images centred on a single LRG after $N$ images are combined into a stack. From left to right top to bottom $N$=1, 10, 1000, 10 000, and 100 000. The bottom right panel shows the 100 000 stack image residual after a model based on the radial average is subtracted from the N=100 000 stack image (bottom left). }
\label{fig:symmetry}
\end{figure}

To demonstrate this, we created a stack centred around a single LRG and looked at the stack after 1, 10, $1\, 000$, $10\, 000$, and $100\, 000$ images are included in the stack. This is shown in Fig.~\ref{fig:symmetry}. The single image does not look symmetric at all. But after only 10 images are included in the stack it is starting to look like random noise, even more so after $1\, 000$ images. By $10\, 000$ images the signal from the LRG (or cluster) is becoming detectable. After $100\, 000$ images the LRG cluster signal is clear and if a model is generated by simply computing the radial average at each pixelised radial bin and subtracted, then the residual looks like the random background noise. 

The LRG catalog uses photometric redshifts, which can have large uncertainties. With a pair of LRGs a moderate redshift uncertainty can result in a large change in the calculated distance between them. To test for the effect of redshift uncertainty we created a subsample of the physically nearby LRG pairs such that both LRGs in the pair had $\delta z \leq 0.01$. This resulted in $N_{\rm LRG}= 310\, 512$. The stacking tests were repeated. The filament excess was the same to within the uncertainty determined from the resampling tests.

In the following subsections we consider different physical causes of the detected excess. 

\begin{table*}
\centering
 \setlength{\tabcolsep}{3.9pt}
\caption{Average values for the filament region in the residual stacking maps. The $\sigma_{\rm stack}$ come from the average noise values of the map divided by $\sqrt{N_{\rm LRG}}$. The uncertainties listed on the measured values are derived from the widths of the resampling distributions. The flux densities listed are using the beam sizes for the each of the different frequencies and or instruments which can be found in Table~\ref{tab:raddat}. }
\label{tab:restab}
\begin{tabular}{lrrrrr}
\hline
\multicolumn{1}{c}{Name} &\multicolumn{1}{c}{$\sigma_{\rm stack}$} & \multicolumn{1}{c}{$\langle T_{\rm fil}\rangle_{\rm physical}$} & \multicolumn{1}{c}{$\langle T_{\rm fil}\rangle_{\rm random}$} &  \multicolumn{1}{c}{$\langle S_{\rm fil}\rangle_{\rm physical}$}& \multicolumn{1}{c}{$\langle S_{\rm fil}\rangle_{\rm random}$} \\
  &  \multicolumn{1}{c}{\scriptsize [K]}& \multicolumn{1}{c}{\scriptsize [K]}&  \multicolumn{1}{c}{\scriptsize [K]} &  \multicolumn{1}{c}{\scriptsize [mJy beam$^{-1}$]} & \multicolumn{1}{c}{\scriptsize [mJy beam$^{-1}$]} \\
 \hline
GLEAM Blue & $0.03$&$0.10\pm0.04$ & $0.01\pm0.04$ & $0.06 \pm .02$ &$-0.01 \pm 0.02$   \\
GLEAM Green &$0.06$ &$0.22\pm 0.06$ & $-0.09\pm 0.05$ & $0.12 \pm 0.03$ &$-0.05 \pm 0.03$  \\
GLEAM Red &$0.13$ &$0.44\pm0.09$ & $-0.06 \pm 0.07$& $0.25 \pm 0.05 $ & $-0.04 \pm 0.04$  \\
OVRO LWA &$0.25$ &$1.1 \pm 0.2$ & $0.1 \pm 0.2$ &$4.0\pm 0.9$ & $0.6 \pm 0.9$  \\
\hline
 & \multicolumn{1}{c}{\footnotesize [Counts s$^{-1}$ arcmin$^{-2}$]}  & \multicolumn{1}{c}{\footnotesize [Counts s$^{-1}$ arcmin$^{-2}$]$_{\rm physical}$}&  \multicolumn{1}{c}{\footnotesize [Counts s$^{-1}$ arcmin$^{-2}$]$_{\rm random}$} & & \\ 
 \hline
 ROSAT Total& $7.1\times 10^{-7}$ & $1.15\times10^{-5} \pm 1.4 \times 10^{-6}$ & $2\times10^{-7} \pm 1.2\times10^{-6}$ & & \\
  ROSAT Soft& $1.8\times 10^{-6}$ & $4.4\times10^{-6} \pm 1.6 \times10^{-6}$ & $-2\times10^{-7} \pm 1.6 \times10^{-6}$ & & \\
   ROSAT Hard& $1.8\times 10^{-6}$ & $9.4\times 10^{-6} \pm 1.8 \times10^{-6}$ & $3\times 10^{-7} \pm 1.8 \times10^{-6}$ & & \\
 \hline
\end{tabular}
\end{table*}

\subsection{Source Contribution}
\label{sec:sources}

We first consider the possibility that the signal seen here is coming from un-subtracted, partially coherent point sources. Faint sources will be confused and below the noise limit and impossible to be subtracted from the maps. However, the excess filament signal is not seen in the control sample, while the control sample too contains faint point sources. Therefore, for the signal seen in the physical pairs to be attributed to galaxies it would have to be galaxies within the filaments. This then demands that the number density of sources along, or within, filaments must be significantly higher than the background field of galaxies for them to add coherently in the stack of the physical pairs. 

It is known from SDSS filament finding catalogs that filaments have an average galaxy over density in the optical of $\sim 5$ \citep{Martinez16,Tempel14}. However, this does not directly translate to the same factor in bright, or detected, radio (or X-ray) sources. Recent work looking at AGN galaxies in the COSMOS field \citep{Schinnerer07,Vardoulaki20} did not find a significant difference in galaxy density when comparing field AGN with those in clusters or filaments. Therefore, for the detected signal to be caused by an over-density of radio galaxies inside filaments, would mean it would have to be fainter sources (below those detected in the deep COSMOS data) and those fainter sources would have have to have much higher numbers in filaments, both in comparison to brighter galaxies and fainter field galaxies. While this possibility cannot be fully excluded, there are no physical or observational reasons to support it at the moment. 

In \citet{Govoni19}, where an intercluster bridge was detected, stacking at the position of optical galaxies in the bridge region, the authors looked at the average source brightness compared to the surface brightness of the bridge and determined a much higher density of sources would be needed to be the cause of the apparent diffuse bridge. Using the average source brightness measured in that work and the source density in the detected bridge, scaled to our average filament size and redshift, yields {\it upper limits} on the source contribution to our signal of $80 - 90\,$per cent. However, it is very likely that the source density from that system is over estimated and/or higher than that for pairs with larger separations or lower mass systems. But, as stated, the source density along filaments is not well constrained. 

Future deep radio surveys such as the MeerKAT International GigaHertz Tiered Extragalactic Exploration survey \citep[MIGHTEE,][]{Jarvis16} or the Square Kilometre Array (SKA) surveys will detect vast numbers of new sources to fainter limits with noise levels projected $\leq 1\, \mu$Jy beam$^{-1}$, and should be able to better constrain the clustering of faint sources.  

Additionally, we can consider the spectral nature of the signal. The spectral index found for this emission is $\alpha=-1.0$. This is consistent with the typical parameters of cosmological strong accretion shocks, whereas the average spectral index for emission from galaxies is $\alpha \sim -0.7$ \citep{Condon92} (although the lobes of AGN can exhibit steeper spectra). Recent work suggests that starburst galaxies may actually show a shallower spectral index at low frequencies ($\nu \lesssim 500\,$MHz) as a consequence of free-free absorption \citep{Galvin18}. Such a strong signal detected in the X-ray would argue that if it was coming from discrete objects it would most likely be active galactic nuclei (AGN) rather than star-forming galaxies. However, low-frequency surveys would argue that the majority of fainter sources that would make up the signal in this flux density range would instead predominantly be star-forming galaxies \citep{Franzen19}.

One way of testing for the presence of unresolved galaxies is to consider the far-infrared to radio correlation \citep[FIRC,][]{deJong85,Condon91,Bourne11}. This relationship is described by the co-efficient $q=F_{\rm FIR}/F_{\rm radio}$ and is typically between 10 to 150 for low radio frequencies \citep{Wong16,Read18}, depending on AGN vs star-forming galaxies (with the lower estimates for AGN).

We use several infrared maps including a $3\, \mu$m ALLWISE map \citep{Wright10,Mainzer11,Lang14}, $60\, \mu$m and 100$\mu$m reprocessed IRAS maps \citep{Miville05}, and a $140\, \mu$m AKARI map \citep{Doi15}. \footnote{The ALLWISE data was at a much higher resolution. The postage stamps were obtained from \url{http://unwise.me} and convolved to 2 arcminute resolution before being combined into a Healpix format} Using the same sample and procedure as for the radio and X-ray stacking, all of the IR stack maps showed a negative average signal in the filament region, consistent with a non-detection. Using the $3\sigma$ sensitivity values of the stacked IR maps the most stringent limit is $q=3\sigma_{\rm IR}/S_{\rm radio} = 6$, from the IRAS $60\, \mu$m and OVRO-LWA maps (with the other maps giving only slightly higher limits). Additionally, stacking this same sample on the {\it PLANCK} tSZ map yields a similar result as seen in previous work of \citet{Tanimura19} and \citet{deGraaff19}, and excess tSZ signal is not expected from galaxies.

The absence of IR emission is an argument against the signal coming from, or being dominated by,  galaxies. 
On the other hand, there is plenty of evidence for diffuse radio emission associated with clusters, such as relics or radio haloes, that has no correlation with   IR emission. While it is possible some of the signal could be coming from such sources, cluster haloes tend to be centred around the cluster mass centre and decrease with brightness away from the centre, and thus any excess from diffuse emission from haloes would not be seen solely between the two LRGs. Also, the majority of detected haloes have reported largest linear sizes of 1 to $3\,$Mpc \citep{Vanweeren19}, whereas the region in which we have detected a residual signal is at an average distance of $\sim 5\,$Mpc from either cluster centre. Similarly with relics associated with clusters, the largest protected distance from the cluster centre of known relics is only $\sim 3\,$Mpc, with an average closer to $0.5$-$1\,$Mpc \citep{Feretti12}. Both the radio power of cluster haloes and relics are usually found to correlate with the X-ray luminosity of a cluster, however, relics do not tend to have spatially correlated X-ray emission and can be found in X-ray poor clusters. Thus while one could argue the radio detection is coming from cluster relics at large distances from the cluster centre, that would not necessarily explain the X-ray results. Numerical simulations also support the fact that peripheral cluster regions and filaments should be largely dominated by non-thermal radio emission, owing to the low particle density and temperature there \citep[e.g.][]{Vazza19}. 

We proceed from here with a discussion of the detected signal under the assumption it is not caused or dominated by emission from galaxies or emission related to either of the clusters. 

\subsection{Magnetic Fields}
\label{sec:bfield}

In this section we investigate the expected magnetic field strength implied by the detected signal, under the assumption that the emission is coming from intergalactic magnetic fields found in filaments. Estimating the magnetic field strength, $B$, in these regions is challenging, but particularly important for studies of the origin of cosmic magnetism, since the dynamics of such low density and relatively relaxed regions should preserve the memory of seed magnetic fields. In order to estimate the magnetic field strength from these measurements we assume equipartition between magnetic field and cosmic ray energy densities \citep{Pacholczyk70,Beck05}. For a synchrotron radio source, the equipartition magnetic field (in gauss) is
\small
\begin{equation}
B_{\rm eq}=\left [ \frac{4\pi (1-2\alpha)(K_0+1)E_{\rm p}^{1+2\alpha}(\nu/2c_1)^{-\alpha} I_{\nu} (1+z)^{3-\alpha}}{(-2\alpha-1)\,c_2(-\alpha)\, l \, \eta \, c_4(i)}\right ]^{1/(3-\alpha)}.
\label{eq:bme}
\end{equation}
\normalsize
Here the volume filling factor of the emitting region(s) is $\eta$, $K_0$ is the ratio of number densities of cosmic ray protons and electrons per particle energy interval within the energy range traced by the synchrotron emission (rather than the commonly used ratio of energy in the protons to that in the electrons), $z$ is the (average) redshift, $l$ is the path length through the source in the line of sight, $\alpha$ is the spectral index ($S(\nu) \propto \nu^{\alpha}$), and $E_{\rm p}$ is the proton rest energy. The synchrotron intensity of the region at the frequency $\nu$ is $I_{\nu}$, where we can use the surface brightness of the filament regions, $S_{\rm fil}$ converted from Jy beam$^{-1}$ to ${\rm erg} \, {\rm s}^{-1} \,  {\rm cm}^{-2} \, {\rm Hz}^{-1} \,{\rm sr}^{-1}$. The constants $c_1$, $c_2$, and $c_4$ are described in Appendix A of \citet{Beck05}.  For strong synchrotron loss (steep cosmic ray energy spectra) the extrapolated low-energy part of the cosmic ray spectrum is overestimated, and so is the resulting field strength from the equipartition equation. It is estimated that for spectra steeper than $\alpha \lesssim -1.1$ that the magnetic field estimates can be significantly overestimated and thus those found in this work, with $\alpha= -1.0$, may be considered upper limits. 

Assuming equipartition, we thus estimate the average magnetic field strength of filaments in the stacked synchrotron detection to be $\sim 60\,$nG, assuming the ratio of number densities of cosmic ray protons and electrons per particle energy interval within the energy range traced by the synchrotron emission to be 100 \citep[or the average value given for both Fermi shock acceleration and Secondary electron CR injection mechanisms,][]{Beck05}, an average redshift $\langle z \rangle = 0.14$, the spectral index $\alpha=-1.0$, a volume filling factor of $0.1$ and the line of sight filament width of 2$\,$Mpc.

For an ultra-relativistic population of cosmic-ray electrons with a power-law distribution of energies, $N_e(E) \propto E^{-n}$, the synchrotron photons also follow a power-law frequency spectrum with a spectral index, $\alpha = -(n -1)/2$. The ratio of the X-ray flux density from inverse-Compton (IC) emission at $\nu_c$ to the synchrotron radio flux density at $\nu_s$ is dependent on the magnetic field strength and the spectral index.
The magnetic field can be derived by looking at the ratio of the flux density from IC emission, $F_c$, at frequency $\nu_c$ to the synchrotron flux density, $F_s$, at frequency $\nu_s$. This ratio is given by
\begin{equation}
\begin{split}
\frac{F_c}{F_s} \, = \, 2.47 \times 10^{19} \, (5.23 \times 10^{3})^{-\alpha} \left ( \frac{T}{1 \rm K} \right )^{3 -\alpha} \\
\frac{b(n)}{a(n)} \left ( \frac{B}{1 \rm G} \right )^{-(1-\alpha)} \left ( \frac{\nu_c}{\nu_s} \right )^{\alpha} \, .
\end{split}
\end{equation}
Here  $T$ is the CMB temperature at the redshift of the filament, the constants $a(n)$ and $b(n)$ have been tabulated in previous works \citep{Ginzburg65,Tucker77}, and $n = 1-2\alpha $.  If we take the radio flux density at $118\,$MHz converted from Jy beam$^{-1}$ to erg s${-1}$ cm$^{-2}$ arcmin$^{-2}$, a spectral index of $\alpha=-1$ and the flux of the combined X-ray bands (in erg s$^{-1}$ cm$^{-2}$ Hz$^{-1}$ arcmin$^{-2}$, assuming all the X-ray is from IC), then the predicted X-ray emission from IC matches the observed X-ray flux for a magnetic field strength of $\sim 30$-$50\,$nG, depending on which X-ray flux is assumed. Attributing a lower contribution of IC to the X-ray flux requires larger magnetic field values.

The magnetic field inferred from this IC argument is roughly consistent with the equipartition estimate, and both are much higher than those from studies using Faraday rotation measures. Rotation measure studies, which are sensitive to the average magnitude and direction of the line-of-sight magnetic field, put estimates of magnetic fields in filaments at $\sim 4$ to $10\,$nG \citep{Osullivan19,Osullivan20,Vernstrom19}. This implies a degree of regularity in filament magnetic fields of $\sim 5$ to $15\,$per cent, indicating either a significant amount of turbulence, many field reversals along the line of sight, and/or that the filament magnetic fields are aligned parallel to or along the filament (as indeed supported by the recent analysis of cosmological simulations, e.g. \citet{Banfi20,Banfi21}), with the resulting geometric effects leading to an underestimation of the line-of-sight magnetic field. 

\subsection{Previous Detections}
\label{sec:previous}

There have been two direct imaging detections of intercluster bridges, by \citet{Govoni19} and \citet{Botteon20}. Both of these used the LOFAR telescope at $140\,$MHz, with \citet{Botteon20} also including LOFAR data at $53\,$MHz. The system in \citet{Botteon20} is at a redshift of $z=0.279$ with a separation between clusters of $\sim 2\,$Mpc. The \citet{Govoni19} system is at a lower redshift of $z=0.07$ with a cluster separation of $\sim 3\,$Mpc. With the use of the two LOFAR bands in \citet{Botteon20} a spectral index was measured with $\alpha=-1.65\pm 0.27$, similar to, although slightly steeper than, the measured index found in this work. 

Both works use a cylindrical volume to estimate the average emissivity of the bridges, $\langle J \rangle$. Govoni et al. found an emissivity of $\langle J_{140 \, {\rm MHz}} \rangle = 8.6 \times 10^{-43}\,$erg s$^{-1}$ Hz$^{-1}$ cm$^{-3}$, while Botteon et al found $\langle J_{144 \, {\rm MHz}} \rangle = 4.02 \times 10^{-43} \,$erg s$^{-1}$ Hz$^{-1}$ cm$^{-3}$. If we also assume a cylindrical volume for the filament region, using a derivation seen in \citet{Murgia09}, with a depth $D=2\,$Mpc we find $\langle J_{154 \, {\rm MHz}} \rangle = 3.2 \times 10^{-45} \,$erg s$^{-1}$ Hz$^{-1}$ cm$^{-3}$. 

The emissivity from the detection in this work is roughly two orders of magnitude below those from the direct imaging detections. Those two detections are themselves roughly an order of magnitude or more below the emissivities of radio cluster haloes \citep[see Fig. S2 of ][for a plot comparing radio cluster halo and filament emissivities]{Govoni19}. The large difference in emissivities between the detected bridges and those in this work implies that the detected bridges are either i) the brightest tail of the distribution of filaments detected here, or ii) as these bridges are between closely separated clusters there is some possibility of more contamination to the bridge region from cluster emission. 

\subsection{Cosmological Simulations}
\label{sec:sims}

\begin{figure*}
\centering
\includegraphics[scale=0.35]{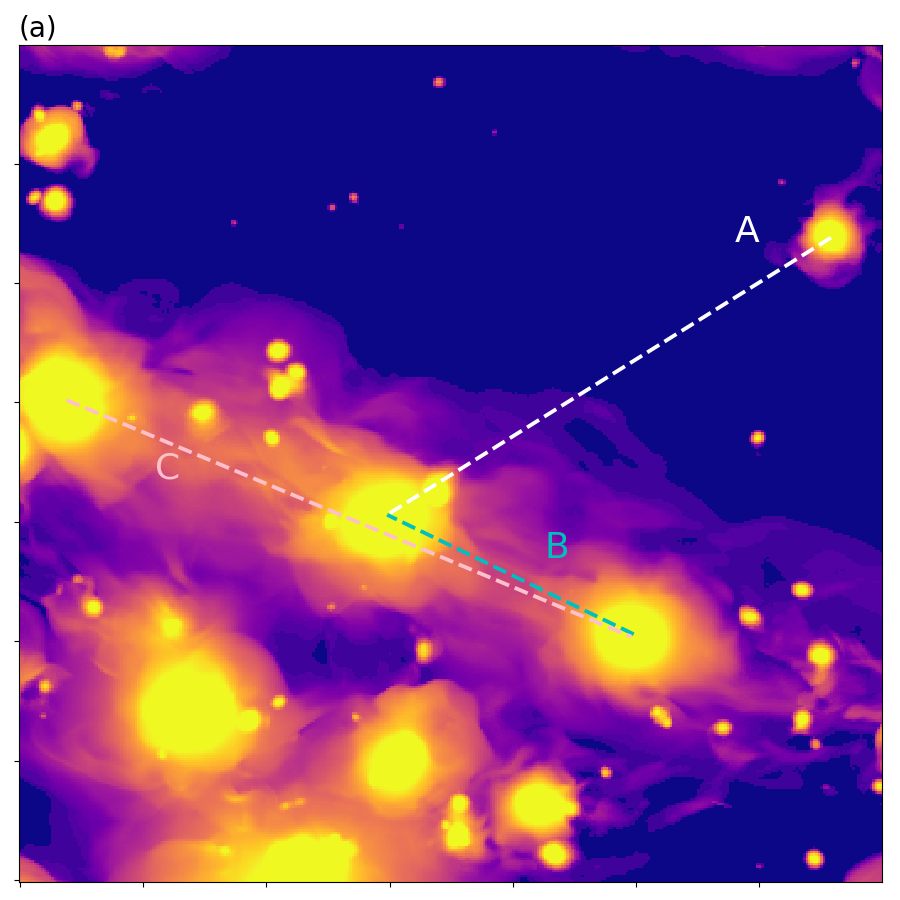}\includegraphics[scale=0.35]{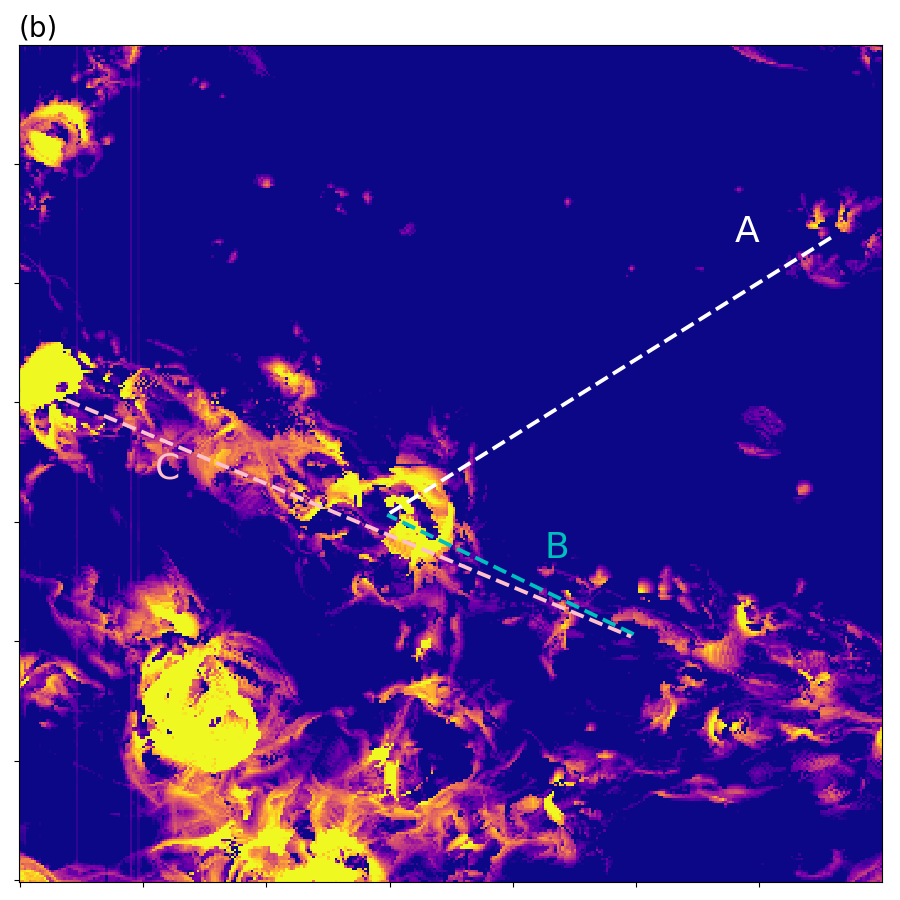}
\includegraphics[scale=0.35]{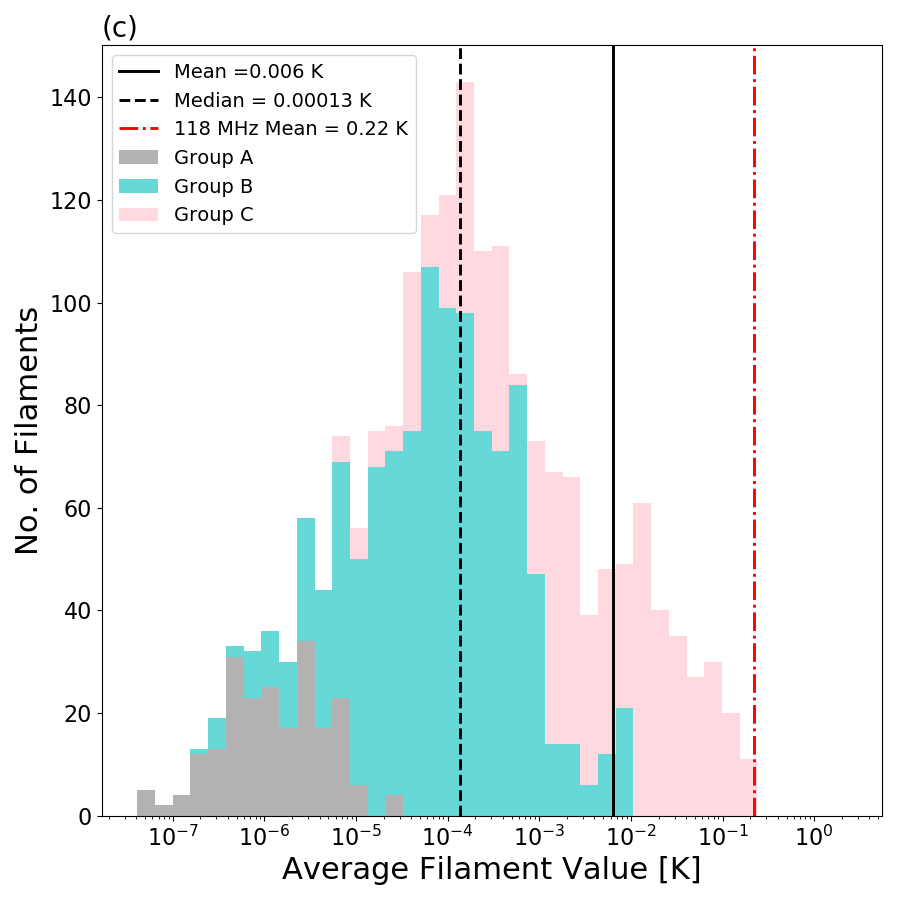}\includegraphics[scale=0.35]{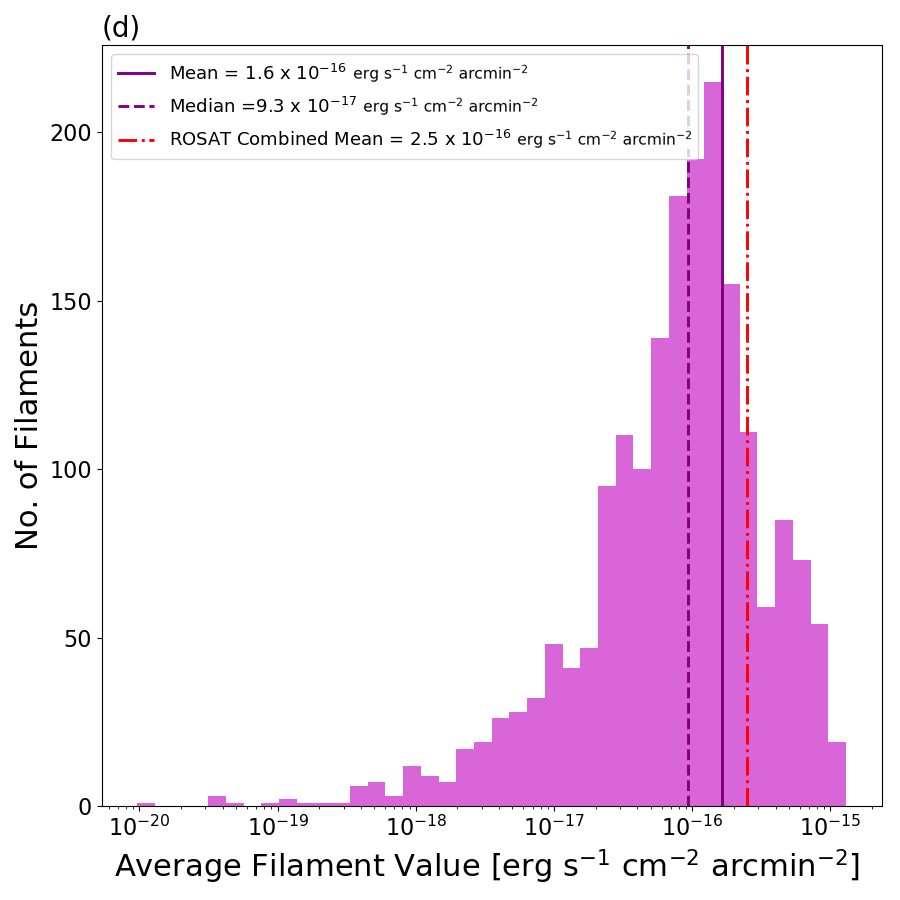}
\caption{ Simulation images and simulated stacking of filaments. Panel (a) shows a section of the simulated X-ray flux map while panel (b) shows the same section with synchrotron flux from the simulation. The lines and labelled components correspond to pairs where: (A) no filamentary emission is between the haloes; (B) there is filamentary emission between the haloes; and (C) there is filamentary emission as well as additional haloes between the haloes pairs. Panel (c) shows the stacked distributions of average radio emission within a box between the haloes, of the same size used in the data stacking, for 2036 pairs of haloes in the simulation (broken up by the groups of A, B, and C), while panel (d) shows the X-ray flux for the same pairs. }
\label{fig:sims}
\end{figure*}

To investigate the nature of the radio and X-ray detections assuming a non-point source, or diffuse, origin, we compare our observed results with thermal and magnetohydrodynamic (MHD) cosmological simulations as detailed in \citet{Vazza19}. The suite of simulations from which we searched and selected mock filaments between cluster pairs has been produced using the cosmological MHD code ENZO \footnote{www.enzo-project.org}. 
The simulation consists of a cubic $100^3\,$Mpc$^3$ comoving volume sampled by $2400^3$ cells,  providing a uniform resolution of 41.6 kpc, and representing the largest MHD simulation in cosmology. The constant dark matter mass resolution is  $m_{dm}=8.62\times 10^6\,M_\odot$. Magnetic fields were initialised at $z=40$ to be $B_0=0.1$~nG (comoving) in all directions, and have been evolved in time  
using the Dedner MHD formalism \citep{Dedner02}. The simulation does not include any additional input of magnetic fields by AGN- or stellar-feedback.  Due to its constant resolution of 40 kpc, the simulation resolves shocks very well across the entire cosmic web, while it under-resolves small-scale dynamo amplification on $<$Mpc scales. 
The diffuse synchrotron emission from the simulated cosmic web was computed in post-processing, by assuming that diffuse shock acceleration (DSA) is the main driver of the synchrotron emission by acceleration of electrons at thermal energies to   relativistic energies in the filaments. Shocks were identified with a velocity-jump based algorithm and the  resulting synchrotron emissivity in the shock downstream was computed based on the formalism by \citet{Hoeft07}.  In their model the synchrotron emissivity is proportional to the fraction of the shock kinetic energy converted into relativistic cosmic rays $\psi_e(\mathcal{M})$ which is in turn computed from the Mach number $(\mathcal{M})$. The normalisation of the emission also depends on the ratio between electrons and protons relativistic energy, assumed to be constant $\xi'_e$ and on the square of the magnetic field strength $B$ at the shock: $f_B=\xi’_e \cdot \psi_e(\mathcal{M} \cdot B^2)$. 

Given the distribution of Mach numbers in the the filaments environment dominated by $\mathcal{M}>5$ shocks, the dependence of the particles energy distribution and in turn of the synchrotron emissivity $I(\nu)$ on the Mach number saturates and the electron acceleration efficiency $\xi_e=\xi'_e \psi_e(\mathcal{M})\simeq 10^{-2}$ is about constant, thus $I(\nu) \propto \xi_e B^2 \nu^{-2}$.

A catalog of mass haloes was generated from the simulation, which can be used directly although in the data haloes are traced by LRGs. The simulations were scaled to the resolution and frequency of the Green GLEAM band (118 MHz), and placed at a redshift of $z= 0.14$. From the main catalog of clusters identified in the simulated volume at $z=0.14$, we further selected pairs of cluster with individual masses $M_{500}>10^{13}M_\odot$, angular distance within $82\, $arcmin$\, \pm 25\,$per cent ,  and with a real linear separation within $10\,$Mpc $\pm 25\,$per cent. This selected a sample of N=2036 cluster pairs with properties similar to the observed sample. Each pair was then aligned to the same normalised grid as was used with the data stacking and rescaled in size (by ensuring conservation of the average surface brightness of each pair). 

In the simulations the cluster pairs can be categorised in 3 types: A) the clusters are near each other but unassociated, with no connecting structure or intervening emission; B) the pairs are connected along filamentary structure with emission peaks between them; or C) the pairs are connected by filamentary structure and also have other haloes or clusters intervening as well. An example of each of these cases is shown in Fig.~\ref{fig:sims}. Case A, where the clusters are just coincidentally nearby each other, should contribute no (or very minimal) signal, and will decrease any average stack signal. On the other hand, case C, with additional masses in-between the pairs, will likely yield a higher signal than if just filamentary emission was present. 

These three scenarios are also expected to be present in the real sample of LRG pairs, and while we cannot distinguish between cases A and B, we are able to identify the pairs which have additional LRGs in-between at approximately the same redshift. When excluding pairs with additional intervening LRGs the stack signal is decreased  by about $60\,$per cent but still amounts to a small detection. While pairs with intervening LRGs may contain some contribution from these LRGs (or clusters), rather than pure filamentary emission, we do not conclude that this is responsible for the detected signal. The control sample LRG pairs should also have intervening LRGs, yet results in no detected signal. We conclude that the signal strength increasing when restricting the physical LRG pairs to those with interveners means that those pairs more likely to actually be connected along larger filamentary structures.

The average filament emission measured for all the simulated pairs at $118\,$MHz is $6\,$mK. The distribution is highly skewed however, with a median value of $0.1\,$mK, suggesting a large impact from outliers. If we look at the distribution of surface brightness values in the box shown in Fig.~\ref{fig:realim} for each image in the stack of the physical LRG pairs, they are also skewed, spanning roughly 6 orders of magnitude but with both positive and negative tails. The average value from the simulations is still about 37 times smaller than the GLEAM Green 118$\,$MHz value that we measure with the stacked data, see panel (c) of Fig.~\ref{fig:sims}. 

These experiments suggest that the observed level of the stacked signal detected in the radio data is not explained by the average population of shocks in simulated filaments, since only a few ($\leq 1\,$per cent) simulated objects can reach such a high emission level. If the simulated intensity was rescaled (by roughly a factor of 6) such that the average value matches the data, a renormalisation of the magnetic field in simulated filaments must be introduced by a factor of 2. For this model, this would imply a 0.6 nG primordial magnetic field and $\langle B \rangle \sim 18\,$nG for the average filament population and $\langle B \rangle \sim 60\,$nG for the more extreme objects of the distribution. This is well within the present constraints on primordial magnetic fields from CMB analysis \citep[$B_0 \leq 4 \rm nG$,][]{Planck16xix}. 

After this rescaling $\simeq 4 \,$per cent of simulated filaments have surface brightnesses in the mJy range with an approximate $3\,$arcminute resolution. If we take the $1.5\,$arcmin resolution of the Phase II of the MWA\citep{Beardsley19}, an rms of $\sim 50\, \mu$Jy beam$^{-1}$ would be necessary for $3\sigma$ direct detections (an order of magnitude lower than estimates for the next deep MWA surveys). For the high-band antenna of the LOFAR telescope, which has much higher resolution, an rms of $\sim10\, \mu$Jy beam$^{-1}$ would be necessary for a resolution of $30\,$arcsec. These estimates do not take into account additional processing that might be necessary to reduce confusion noise and enhance large-scale emission to enable a direct image. 

Alternatively, these results may hint at a higher acceleration efficiency of relativistic electrons by strong structure
formation shocks than assumed. In order to match the 118MHz stacked signal, the maximum acceleration efficiency
should be increased from  1 to $\sim 30$ per cent in shocks with
Mach numbers $\geq 5$. While there are large uncertainties in the physical picture of particle acceleration of shocks \citep{Bykov19} and the latest results from particle in cell simulations \citep{Xu20}, such large acceleration efficiency appear implausible.  A higher acceleration efficiency may be mimicked by the additional contribution from older accelerated relativistic electrons, which shocks in filaments 
are re-energising \citep[e.g.][]{Govoni19}. In summary, a $~37 $ times higher value of  $\xi_e B^2 $ than what is assumed in the simulation appears required to explain the stacking detection using only shock
acceleration. This renormalisation appears physically justified by magnetic fields which are significantly higher than what is assumed in the model, combined with a slightly higher acceleration efficiency of electrons by strong shocks. 

We also repeated the above procedure for a different set of simulations, described in \citet{Gheller20}. These simulations are lower resolution (83 kpc pixel size, 85$^3$ Mpc$^3$ total volume) but include different physical scenarios such as :  an astrophysical scenario with $B$ injected by AGN and star forming regions, following the cosmic star formation history; a primordial model with a stronger primordial magnetic field of $1\,$nG; and a dynamo model in which at run-time intracluster medium-like amplification of magnetic fields is assumed, included within filaments. The results from the increased primordial field and the dynamo scenario came out on the same order as the high resolution model. The average emission with the astrophysical injection model is about two orders of magnitude lower than all the others ($\sim 0.036\,$mK) as the emission in this model drops off much more drastically outside of clusters. Even with the increased primordial field, the amplification of $B$ inside and around clusters is better with the higher resolution model, whereas the lower resolution increased primordial field model has a higher average signal from the filaments away from cluster regions, yielding similar averages when looking over all the pairs.

The average X-ray surface brightness from the simulations comes out almost exactly the same as the surface brightness found from stacking the real LRG pairs  (depending on which model for the X-ray flux is used), see Fig.~\ref{fig:sims}. The emission from the simulation does not include any inverse-Compton contribution to the X-ray flux. Thus for the signal seen in the data to be consistent with the simulations this implies a small contribution from inverse-Compton emission, which would require an even larger magnetic field strength than the $\sim 40\,$nG found above. We note here that the predicted X-ray in the simulations is sensitive to the assumed metallicity (here the metallicity is 0.3 of solar) as well as the exact energy range (the energy range used in the simulations is 0.3 to 2 keV rather than the 0.1 to 2.4 keV band of ROSAT). The average X-ray surface brightness found here from the real LRG pairs is $\sim 4$-$10\,$times higher than that reported by \citet{Tanimura20b}, who also stacked on the RASS data. \citet{Tanimura20b}  used more ROSAT channels, a different sample of filaments to stack, and a slightly different method of analysis all of which could explain the discrepancy.

\subsection{Dark Matter}
\label{sec:darkm}
A possible explanation often considered for the excess detected in the diffuse radio background is synchrotron emission from dark matter annihilation and decay \citep[e.g.][]{Fornengo11}. While these models generally consider the bulk of the dark matter, and emission, to be coming from haloes (both galactic and those in clusters) it is nevertheless still possible for there to be some contribution coming from filaments. While this scenario has not been previously looked at in detail, we can get an estimate using some simple assumptions.

\begin{figure}
\includegraphics[scale=0.57]{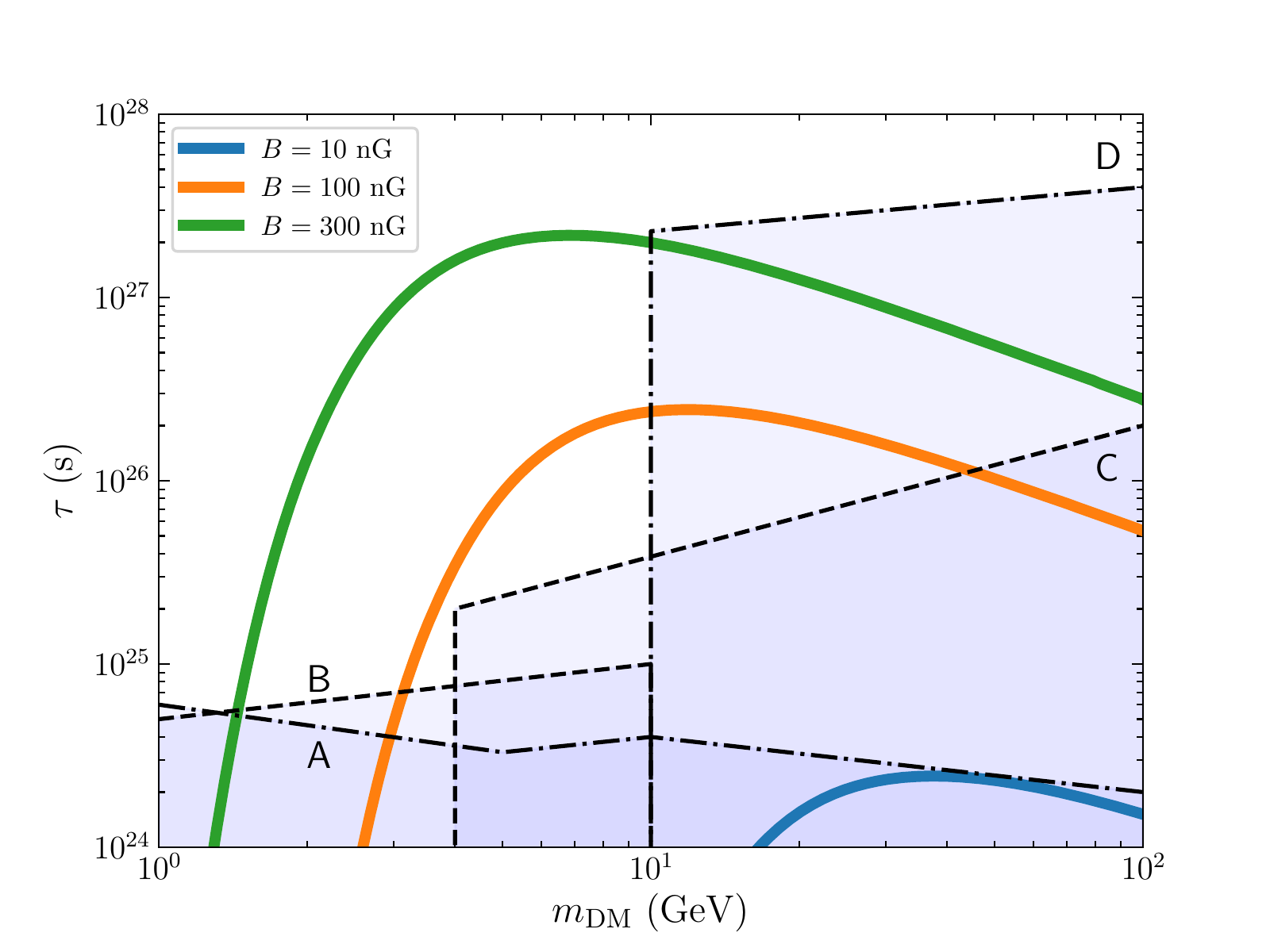}
\caption{ Lower bounds at the 95$\,$per cent confidence level on the dark matter lifetime $\tau$ as a function of the dark matter mass $m_{\rm DM}$, for decay into $e^+ e^-$ and for some representative values of the magentic field $B$. Dashed and dot-dashed lines show the current bounds from \citet{Slatyer:2016qyl} (curve A), \citet{Essig:2013goa} (B), \citet{Massari:2015xea}(C) and \citet{Blanco:2018esa} (D).}
\label{fig:bounds}
\end{figure}

The electron source term for dark matter annihilation is defined in terms of the annihilation cross section $\left\langle\sigma_{a} v\right\rangle$, the dark matter particle mass $m_{\rm DM}$ and the dark matter density $\rho$ as:
\begin{equation}
Q_{e}(E)=\left\langle\sigma_{a} v\right\rangle \frac{\rho^{2}}{2 m_{\chi}^{2}} \times \frac{d N_{e}}{d E}(E) 
\label{eq:source}
\end{equation}
where $d N_{e}/d E$ is the electron spectrum produced in the annihilation event. In case of dark matter decay:
\begin{equation}
Q_{e}(E)= \frac{\rho}{\tau_D m_{\chi}} \times \frac{d N_{e}}{d E}(E) 
\end{equation}
where $\tau_D$ is the particle lifetime. We model the filament as a cylinder with diameter $D = 2$ Mpc,  length $L=8$ Mpc, total mass $M = 4 \times 10^{13} M_\odot$ \citep{Yang20} and of uniform density $\rho = M/(\pi R^2 L) = 1.6 \times 10^{12}  M_\odot {\rm Mpc}^{-3} \simeq 10 \rho_{c}$. Filaments are assumed edge-on. If turbulence is strong enough (we will comment below about this assumption), electrons are confined at their injection site and their equilibrium spectrum is obtained as:
\begin{equation}
n_{e}(E)=\frac{1}{b(E)} \int_{E}^{\infty} d E^{\prime} Q_{e}(E^{\prime})
\end{equation}
where $b(E) \simeq 2.7\times 10^{-17} {\rm GeV} \,{\rm s^{-1}}\, ({E}/{\rm GeV})^2$ is the energy-loss due to IC on CMB (synchrotron losses are subdominant for our assumptions on the magnetic field). The synchrotron emissivity is obtained as: \begin{equation}
j_{\text {sync}}(\nu) = \int d E P_{\text {sync}}(E, \nu) n_{e}(E)
\end{equation}
where $P_{\text {sync}}(E, \nu)$ is the usual radiative synchrotron emission power:
\begin{equation}
P_{\text {sync}}(E, \nu)=\frac{\sqrt{3} e^{3}c}{4\pi\epsilon_0 m_{e} c^{2}} B F\left(\nu / \nu_{c}\right)
\label{eq:power}
\end{equation}
where $\nu_{c} \equiv 3 c^2 e /(4 \pi) /\left(m_{e} c^{2}\right)^{3} B E^{2}$ and $F(t) \equiv t \int_{t}^{\infty} d z K_{5 / 3}(z)$.
For completeness, we assume a constant magnetic field $B=100$ nG and we do not include synchrotron self-absorption in the source.
The intensity at frequency $\nu$ is then:
\begin{equation}
I(\nu) = \frac{1}{\Delta\Omega}\int \frac{d\Omega d s}{4\pi}\, j_{\rm sync}(\nu) \longrightarrow \frac{D}{4\pi} j_{\rm sync}(\nu) 
\end{equation}
where we apply our approximating assumption on the geometry of the filament. As customary, we quote our results in terms of the 
brightness temperature
$T_{b}={I(\nu) c^{2}}/{2 k \nu^{2}}$.

Results are shown in Table \ref{tab:DM} for a selection of the dark matter particle-physics parameters and for a magnetic field $B = 100$ nG. Models A-C refer to decaying dark matter, and the lifetimes are set at (or above by a factor of 3, for Model A) the current bounds \citep{Fornengo:2013xda}. Model D and E refer to an annihilating dark matter with $m_{\rm DM}=100$ GeV and with a canonical thermal cross section, which is close to its current bound \citep{Ackermann:2015zua,Slatyer:2015jla,Ando:2015qda,Calore:2018sdx}.

Since the filament is a significantly less dense structure than a typical galaxy, the annihilating dark matter signal (which is proportional to $\rho^2$) turns out to be suppressed as compared to the decaying dark matter one (proportional to $\rho$) when we sit at the constraints for $\langle \sigma v \rangle$ and $\tau_D$ obtained in galaxies. For annihilation we  obtain brightness temperatures below the $\mu$K level. In our estimate, we assumed a homogenous density: a density profile steeper toward the central axis of the filament or the presence of clumps inside the filament (both options could be considered as reasonable, also based on numerical simulations) could boost the annihilation signal by a factor that depends on the specific details of the mass distribution inside the filament. A boost factor of 100 could be feasible, making the annihilating flux at the level of few to tens of $\mu$K. See Table \ref{tab:DM} for two quantitative examples.

On the contrary, decaying dark matter can provide much larger brightness temperatures. For hadronic annihilation, the strongest bounds on the lifetime are around $\tau_D > 10^{28}$ s \citep{Fornengo:2013xda}, while for the $e^+ e^-$ channel $\tau_D > 2 \times 10^{25}$ s \citep{Essig:2013goa,Massari:2015xea, Slatyer:2016qyl} \citep[a confirmation of the EDGES observations;][]{Bowman:2018yin} could improve the bound to $\tau_D > 3 \times 10^{26}$ s \citep{Mitridate:2018iag} for masses below 10 GeV and $\tau_D > 2 \times 10^{27}$ s for masses above \citep{DiMauro:2014iia, Blanco:2018esa}. Table \ref{tab:DM} shows that in this case brightness temperatures from tens to hundreds of mK can be obtained for the leptonic channel. If the DM mass is below 10 GeV, it turns out to be possible to approach the observed emission level \citep[notice that in this case we adopted a conservative lifetime 3 times larger than the current bound from X-rays,][]{Essig:2013goa}.

Figure~\ref{fig:bounds} instead shows the bounds on the dark matter lifetime obtained from the observed filament temperatures: these bounds are competitive for dark matter masses in the range 3-$10\,$GeV and for magnetic fields in excess of about $50\,$nG.

The results presented above refer to a filament mass $M = 4 \times 10^{13} M_\odot$, for a magnetic field $B=100$ nG and optimal magnetic containment of the electrons in the filament. The decaying and annihilating signals scale as $M B^n$ and $M^2 B^n$, respectively, with $n<2$, depending on the actual electron spectrum. An increase of the filament mass of an order of magnitude would directly reflect this increase in the brightness temperatures shown in Table \ref{tab:DM}, while a reduction of $B$ to 10 nG would reduce the predicted temperature by a factor of at least 50.
Magnetic containment depends on the turbulent component of the magnetic field. Confinement time scales as $\tau_{\rm conf}(E) \sim D^2/2K(E)$, where $K(E) = K_0 (E/{\rm GeV})^{\delta}$ is the diffusion coefficient. By adopting a Kolmogorov spectrum $\delta = 1/3$ and a $K_0$ related to the magnetic field  fluctuations $\delta B$ as $K_0 \propto (B/\delta B)^2 B^{-\delta}$ \citep{Regis:2014koa} and rescaling it to the Milky-Way value $K_0^{\rm MW} = 3\times 10^{28} {\rm cm}^2 {\rm s}^{-1}$ as $K_0^{\rm fil} \sim K_0^{\rm MW} \times k (B_{\rm MW}/B)^\delta$ with $k = (B/\delta B)^2_{\rm fil}/(B/\delta B)^2_{\rm MW}$, we obtain that the confinement time is typically larger than the cooling time $\tau_{\rm cool}(E) = E/b(E)$ at the energies of interest (few GeV) unless $k \gtrsim 10^{4}$. As an example, following the approximation of Section 3.2.4 of \citet{Fornengo:2011iq}, we obtain that the temperatures of Model B get reduced by a factor 0.05 when $k = 5 \times 10^{4}$.

These estimates show that annihilating dark matter would produce filament brightness temperatures at most at the $\mu$K level. Decaying dark matter temperatures can instead reach the 10-100 mK level (depending on the frequency) without conflicting with current bounds on the particle lifetime \citep{Fornengo:2013xda,Essig:2013goa,Massari:2015xea, Slatyer:2016qyl,Blanco:2018esa}. The largest temperatures, close to those observed, are obtained for a dark matter particle with a mass around 5-10 GeV decaying into $e^+e^-$ pairs. Hadronic decays are instead disfavoured. 

\begin{table}
\centering
\begin{tabular}{cccccc}
\hline
 $\nu$ & Model A  & Model B  &  Model C & Model D & Model E \\
(MHz) & (dec) & (dec) & (dec)& (ann)  & (ann)\\
 & (mK) & (mK) & (mK) & ($\mu$K)  & ($\mu$K)\\
\hline
154 & 23 & 11 & 0.1   & 0.07 & 0.01 \\
118 & 110 & 29 & 0.2  & 0.13 & 0.01 \\
88 & 498 & 74 & 0.5   & 0.27& 0.04 \\
73 & 1175 & 131 & 1.0  & 0.44 & 0.07 \\ 
\hline
\end{tabular}
\caption{Filament temperature for decaying and annihilating dark matter. The DM mass and lifetimes refer to: 
Model A: decay into $e^+e^-$, $m_{\rm DM} = 5$ GeV, $\tau_D = 10^{26}$ s; 
Model B: decay into $e^+e^-$, $m_{\rm DM} = 10$ GeV, $\tau_D = 2 \times 10^{27}$ s; 
Model C: decay into $\bar b b$, $m_{\rm DM} = 1$ TeV, $\tau_D = 6 \times 10^{27}$ s; 
Model D: annihilation into $e^+e^-$, $m_{\rm DM} = 100$ GeV, $\langle \sigma v \rangle = 3 \times 10^{-26}$ cm$^3$ s$^{-1}$.
Model E: annihilation into $\bar b b$, $m_{\rm DM} = 100$ GeV, $\langle \sigma v \rangle = 3 \times 10^{-26}$ cm$^3$ s$^{-1}$.
In all cases, $B = 100$ nG.}
\label{tab:DM}
\end{table}

\section{Conclusions}
\label{sec:conclusions}

In this paper we report on the first robust detection of the stacked radio signal from large ($1$-$15\,$Mpc) filaments connecting pairs of luminous red galaxies, and a confirmation of the stacked X-ray filament signal from \citet{Tanimura20b}. This signal appears compatible with the non-thermal synchrotron emission from the shocked cosmic web, and provides a direct evidence for one of the pillars of the physical picture of structure formation in the Universe.

We have shown that the radio signal is not dominated by unsubtracted point sources based on IR stacking results and the typical radio spectral index, with a radio spectral index of the detected signal of $\alpha = -1.0$.  We also investigate possible instrumental or systematic effects as a potential cause, such as beam sidelobes, and exclude these as the cause of the signal.  

 The ROSAT X-ray detections are in agreement with predictions from cosmological thermal simulations, while a factor of roughly 5 higher than the previous X-ray filament stacking results from \citet{Tanimura20b}. Newer X-ray missions, such as eROSITA or the Advanced Telescope for High-ENergy Astrophysics (ATHENA), will be able to probe the X-ray sky to new depths and follow up this result with spectroscopic capabilities in detail.

Assuming a diffuse filamentary origin for the signal, estimates for the magnetic field strength from both equipartition and Inverse Compton arguments are in the range of $30$ -$60\,$nG. Comparing with predictions from simulations of shocked intergalactic gas, the observed radio signal is at 30-40 times higher than predicted. This implies that the seed magnetic field strength would have to be higher than simulated, but still within the bounds from cosmological results. Follow-up work examining this method with more detailed and complex simulations is underway in Hodgson et al., in prep. 

We also investigated the contribution from dark matter annihilation and decay. We find a dark matter with a mass of $5$-$10\,$GeV could produce a signal close to observations if it decays leptonically, although for more general cases the emission is at least one order of magnitude lower. 

While these detections may be dominated by a subset of extreme objects, it nonetheless shows for the first time that magnetised structure and some level of cosmic-ray electron acceleration exists outside of intracluster overlapping regions. These regions may be more highly magnetised than previously believed and/or subject to more efficient shock acceleration, which can be tested by new large-scale and or high-resolution simulations and deep LOFAR surveys and or the Square Kilometre Array (SKA) Low to look for more individual filaments.

\section{Data Availability}
All data used is publicly available. The X-Ray RASS maps, the IRAS, and the AKARI maps can be obtained from \url{http://cade.irap.omp.eu/dokuwiki/doku.php}. The Owens Valley Long Wavelength Array map was obtained from \url{https://lambda.gsfc.nasa.gov/product/foreground/fg_ovrolwa_radio_maps_get.cfm}. The ALLWISE individual cutout images are available from \url{http://unwise.me}, full low resolution healpix mosaics of the ALLWISE/UNWISE maps can be obtained by contacting T.V. Information on obtaining GLEAM images can be found at \url{https://www.mwatelescope.org/gleam}.

\section{Acknowledgments}
The authors would like to acknowledge the help, work, and sharp eye of T. Hodgson in finding and resolving a significant normalisation error of the simulations in the draft version of the paper. The authors would also like to acknowledge the contributions of H. Tanimura and N. Hurley-Walker. We acknowledge the use of data provided by the Centre d'Analyse de Données Etendues (CADE), a service of IRAP-UPS/CNRS (http://cade.irap.omp.eu, \citet{Paradis12})
     F. V.  and N. L. acknowledge financial support from the European Union's Horizon 2020 program under the ERC Starting Grant "MAGCOW", no. 714196. We also acknowledge the usage of online storage tools kindly provided by the INAF Astronomica Archive (IA2) initiave (http://www.ia2.inaf.it). 
 We acknowledge the  usage of computational resources on the Piz Daint supercomputer at CSCS-ETHZ (Lugano, Switzerland) under projects s701 and s805 and at the J\"ulich Supercomputing Centre (JFZ) under project HHH42 and ``stressicm".
    N. F. and E. P. acknowledge funding from: {\sl Departments of Excellence} grant awarded by the Italian Ministry of Education, University and Research ({\sc Miur}); Research grant {\sl The Dark Universe: A Synergic Multimessenger Approach}, No.$\,$2017X7X85K funded by the Italian Ministry of Education, University and Research ({\sc Miur}); Research grant {\sl The Anisotropic Dark Universe}, No.$\,$CSTO161409, funded by Compagnia di Sanpaolo and University of Torino;  Research grant {\sc TAsP} (Theoretical Astroparticle Physics) funded by Istituto Nazionale di Fisica Nucleare ({\sc Infn}). E. P. acknowledges a grant from the Universit\'e Franco-Italienne under Bando Vinci 2020.
    The International Centre for Radio Astronomy Research (ICRAR) is a Joint Venture of Curtin University and The University of Western Australia, funded by the Western Australian State government.
    The Dunlap Institute is funded through an endowment established by the David Dunlap family and the University of Toronto. J.W. acknowledges the support of the Natural Sciences and Engineering Research Council of Canada (NSERC) through grant RGPIN-2015-05948, and of the Canada Research Chairs program.

\bibliographystyle{mnras}

\bsp

\appendix
\section{{Wavelet Model}}
\label{sec:Wavelet Model}

\begin{figure*}
\centering
\includegraphics[scale=0.27]{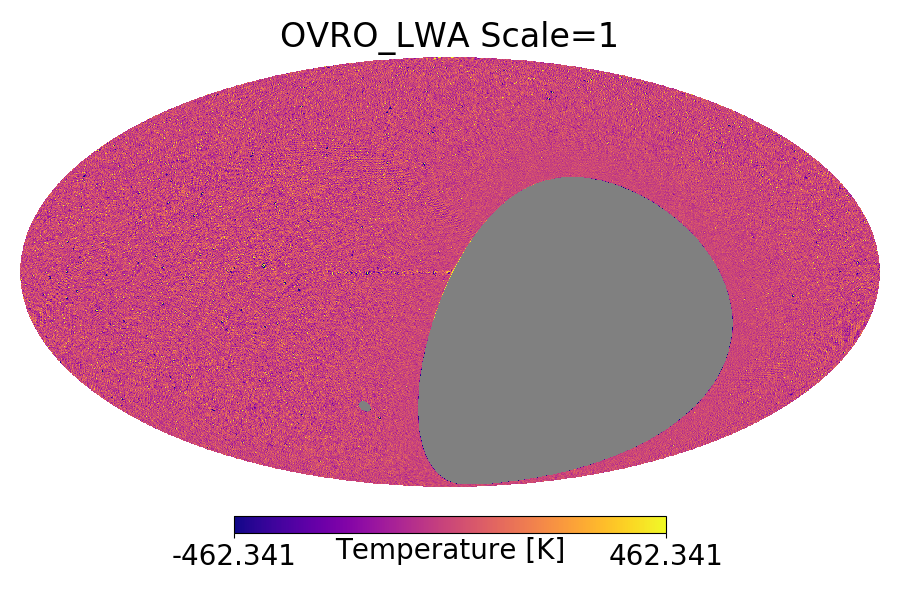}\includegraphics[scale=0.27]{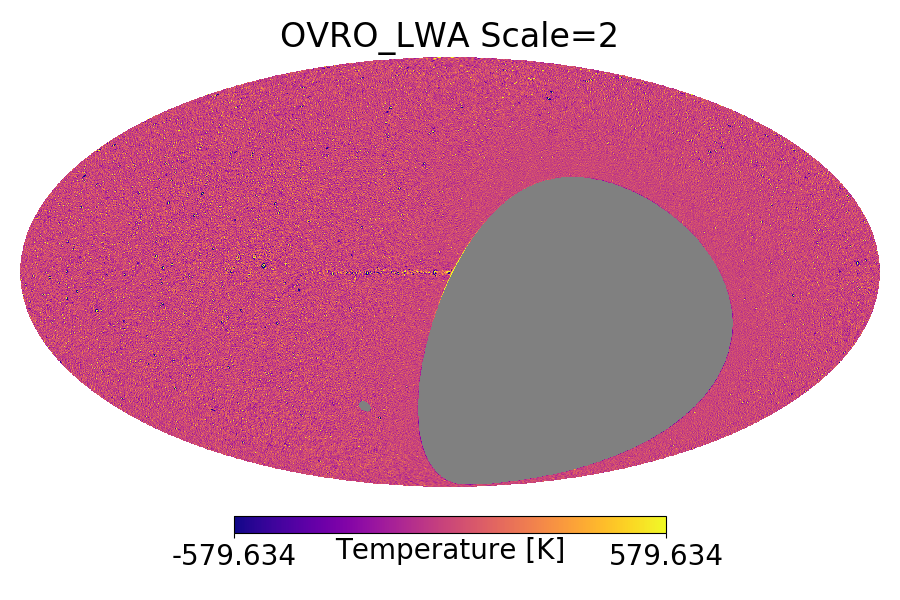}\includegraphics[scale=0.27]{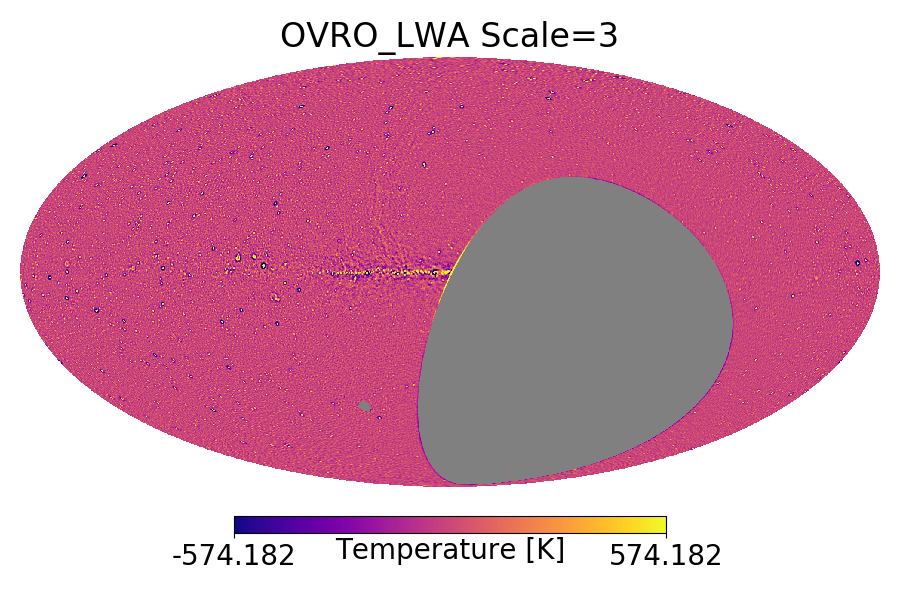}
\includegraphics[scale=0.27]{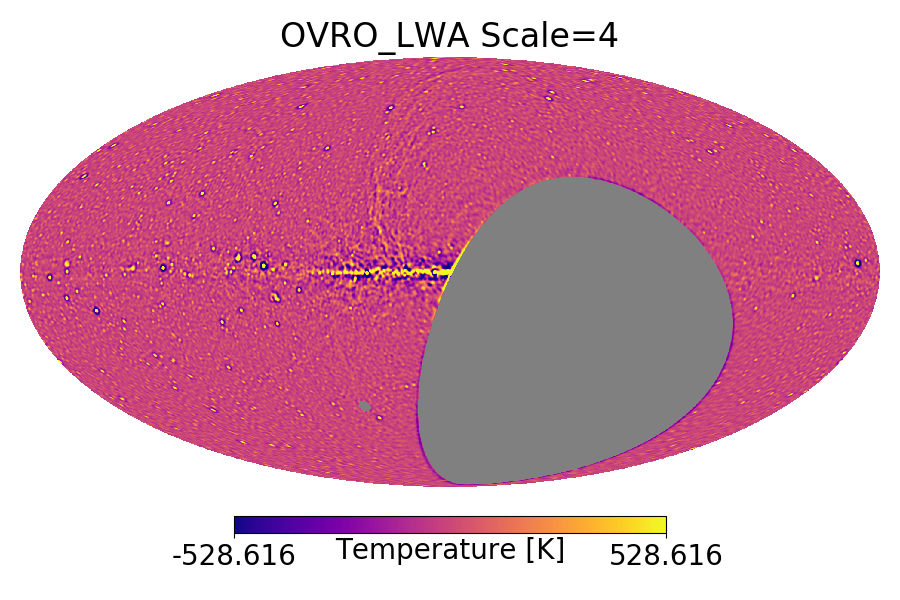}\includegraphics[scale=0.27]{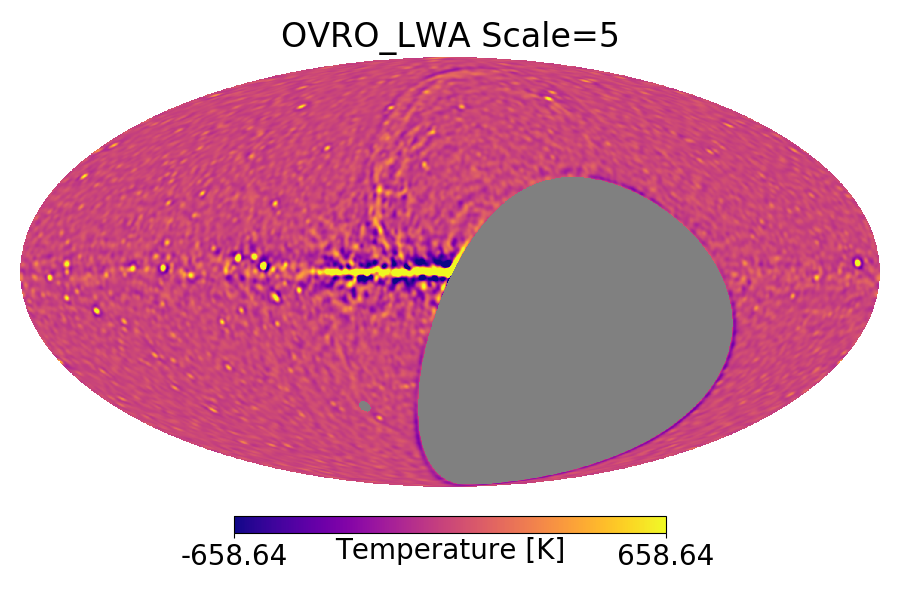}\includegraphics[scale=0.27]{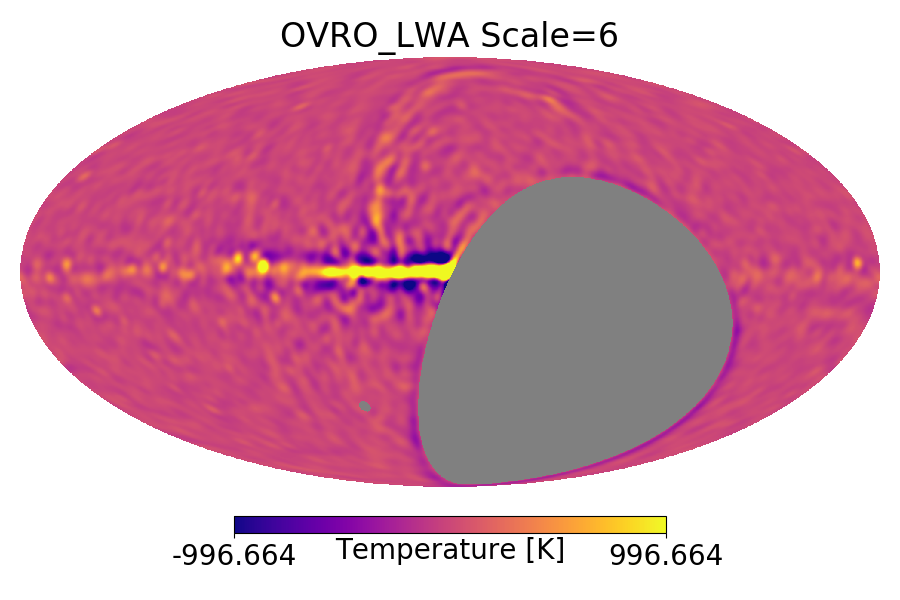}
\includegraphics[scale=0.27]{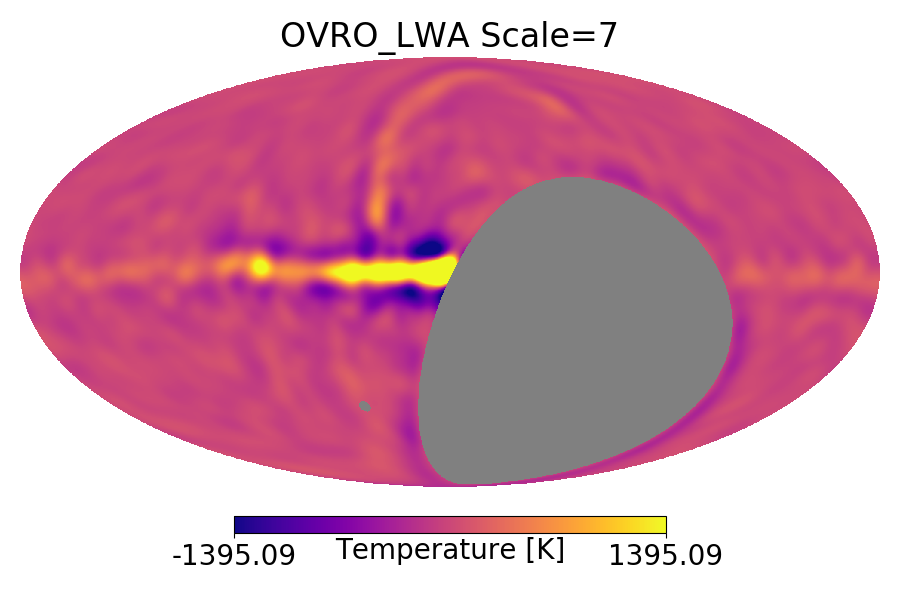}\includegraphics[scale=0.27]{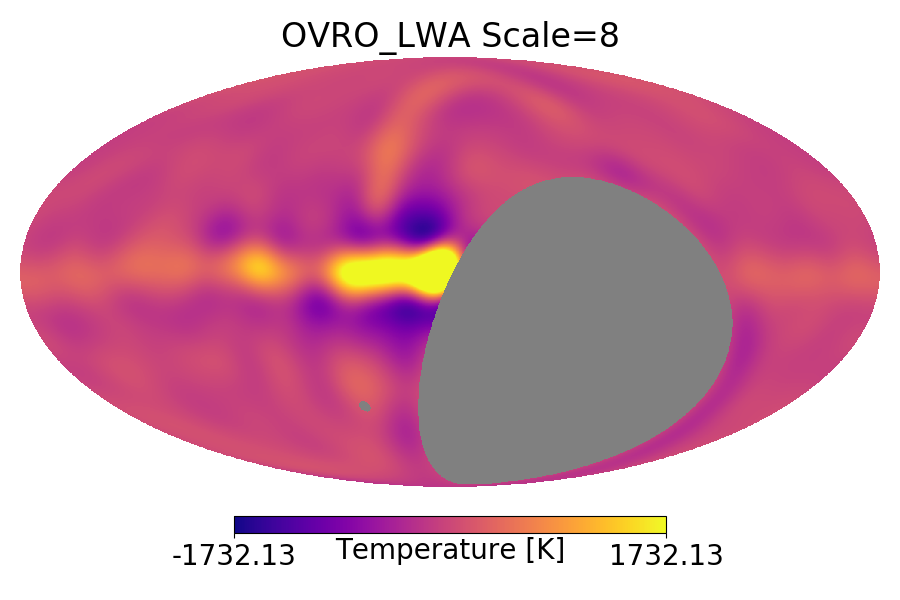}\includegraphics[scale=0.27]{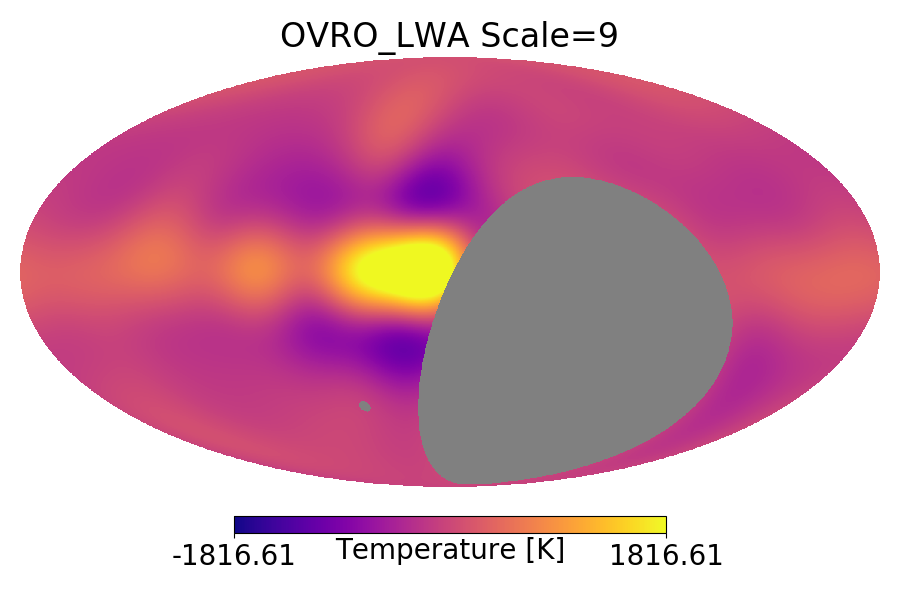}

\caption{The different wavelet scales of the Owens Valley Radio data.}
\label{fig:wave1}
\end{figure*}

\begin{figure*}
\centering
\includegraphics[scale=0.27]{Fig5a.png}\includegraphics[scale=0.27]{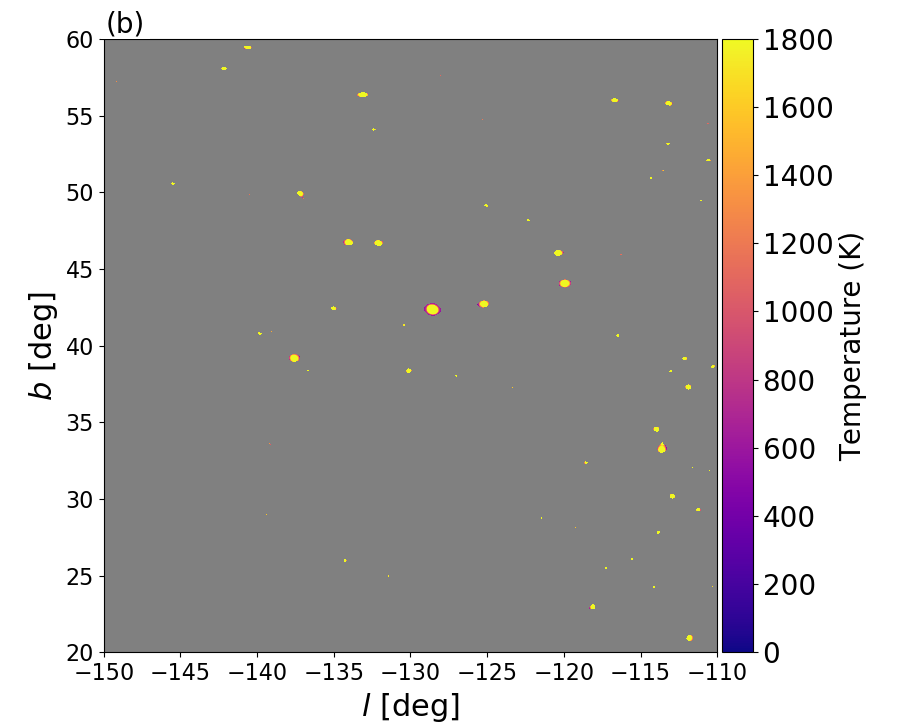}\includegraphics[scale=0.27]{Fig5c.png}

\caption{Zoom in section of the Owens Valley map. Panel (a) shows the map before any subtraction. Panel (b) shows the wavelet based model of the region and panel (c) shows the residual.}
\label{fig:wave2}
\end{figure*}

In order to model and subtract the sources in the images we use the method of wavelet decomposition, or wavelet deconvolution. \citet{Vikhlinin98} gives a description of the process used in analysis of X-ray images. The process uses the software described above in the main text, Interactive Sparse Astronomical Data Analysis Packages (ISAP), and the \textsc{mrs$\_$uwttrans} program to decompose the image into a set of scales. These scales have been created by convolving by a kernel which consists of a positive core and negative outer ring, so that the integral over the kernel is zero. An example of these scales for the Owens Valley map is shown in Figure~\ref{fig:wave1}.

Once decomposed into the different wavelet scales the process for making a model proceeds as follows:
\begin{enumerate}[label=\arabic*)]
    \item Starting with the smallest scale, compute the rms of the scale image;
    \item locate peaks, or pixels, that are $> 5\sigma$ and add them to a blank model;
    \item convolve this model with a Gaussian beam;
    \item subtract this model from the original image;
    \item decompose the residual into the corresponding wavelet scales again;
    \item repeat steps 1-4 for the second smallest scale. 
\end{enumerate}

This process can be continued for larger scales if needed. However, if the goal is to find point source emission, only the smallest scales should be used for generating the model. Figure~\ref{fig:wave2} shows an zoom in of a section from the Owens Valley map with the resulting model from the smallest wavelet scales and the final residual. 

The subtraction is not exact, as can be seen in Fig.~\ref{fig:wave2} and Fig.~\ref{fig:pssub}. The main reason for this is likely the fact that in the modelling process described here, a single-sized circular Gaussian beam is used for the model for the whole sky. However, in all of the radio maps this is not the case and the true beam has a varying shape and size across the sky and therefore the model will not provide a perfect subtraction. It did provide a close approximation for our purposes here, but in an ideal setup one would use the residuals from the initial image deconvolution process.

\label{lastpage}

\typeout{get arXiv to do 4 passes: Label(s) may have changed. Rerun}

\end{document}